\documentclass[letterpaper,english,reprint, aps, nofootinbib]{revtex4-1}
\usepackage[T1]{fontenc}
\usepackage[latin9]{inputenc}
\usepackage{float}
\usepackage{amsmath}
\usepackage{amssymb}
\usepackage{graphicx}

\makeatletter

\pdfpageheight\paperheight
\pdfpagewidth\paperwidth

\makeatother

\usepackage{babel}
\begin{document}

\title{Interaction patterns and diversity in assembled ecological communities}

\author{Guy Bunin}

\affiliation{Department of Physics, Massachusetts Institute of Technology, Cambridge,
Massachusetts 02139, USA}
\begin{abstract}
The assembly of ecological communities from a pool of species is central
to ecology, but the effect of this process on properties of community
interaction networks is still largely unknown. Here, we use a systematic
analytical framework to describe how assembly from a species pool
gives rise to community network properties that differ from those
of the pool: Compared to the pool, the community shows a bias towards
higher carrying capacities, weaker competitive interactions and stronger
beneficial interactions. Moreover, even if interactions between all
pool species are completely random, community networks are more structured,
with correlations between interspecies interactions, and between interactions
and carrying capacities. Nonetheless, we show that these properties
are not sufficient to explain the coexistence of all community species,
and that it is a simple relation between interactions and species
abundances that is responsible for the diversity within a community.
\end{abstract}
\maketitle
\global\long\def\mean{\operatorname{mean}}
\global\long\def\var{\operatorname{var}}
\global\long\def\std{\operatorname{std}}
\global\long\def\corr{\operatorname{corr}}
\global\long\def\cov{\operatorname{cov}}
\global\long\def\sign{\operatorname{sign}}
\global\long\def\erf{\operatorname{erf}}

Networks of species interactions and their structure are central objects
of study in community ecology, both in terms of the organization of
the links and the strength of the interactions. Network structure
has been related to increase in maximal diversity and species abundance
\citep{bastolla_biodiversity_2005,bastolla_architecture_2009} and
to ecosystem functioning \citep{fuhrman_microbial_2009,thompson_food_2012}.
Interaction patterns tell us about underlying mechanisms of interaction
such as competition over resources, and about the evolutionary and
assembly history of the community \citep{may_theoretical_2007}. In
addition, the network structure is shaped by the requirements of stability
\citep{montoya_ecological_2006,mccann_diversitystability_2000,mccann_weak_1998}
both internally and in the face of migration to and from a pool of
available species \citep{macarthur_theory_2015,leibold_metacommunity_2004}.
The resulting abundances must be positive, a requirement known as
feasibility \citep{roberts_stability_1974,bastolla_biodiversity_2005,rohr_structural_2014}.
Such constraints are especially important in conditions where interactions
are a dominant factor \citep{leibold_metacommunity_2004,fisher_transition_2014}.

These constraints imply that viable communities are not arbitrary
collections of species from the pool, but instead have special properties.
This is expected to affect the network properties. For example, species
that suffer from strong competition are less likely to persist, so
weaker competitive interactions might be over-represented in the community,
as was indeed observed in simulations \citep{kokkoris_patterns_1999-1}.
Similarly, species with higher carrying capacities might have better
chances to persist, biasing this distribution with respect to that
of the entire pool. Beyond such considerations, a framework giving
definite, quantitative predictions for these effects, and for the
emergence of more complex patterns has thus far been lacking.

To shed light on this process we turn to community assembly models,
where the interactions between all species in the pool \textendash{}
as would be measured in the local conditions \textendash{} are modeled.
Such models have provided insight into the influence of the assembly
process and the existence of multiple equilibria \citep{gilpin_multiple_1976,diederich_replicators_1989,drake_mechanics_1990,post_community_1983,case_invasion_1990,law_permanence_1996,morton_regional_1997,capitan_statistical_2009,biscari_replica_1995,kessler_generalized_2015,fried_communities_2016},
the resulting species abundance \citep{rieger_solvable_1989,opper_phase_1992,tokita_species_2004,yoshino_rank_2008},
growth of resistance to invasion \citep{case_invasion_1990,morton_regional_1997,capitan_statistical_2009},
the effects of noise and rates of relaxation following a change in
the community \citep{rieger_solvable_1989,opper_phase_1992,fisher_transition_2014,kessler_generalized_2015,fried_communities_2016},
and ecosystem function \citep{goudard_nontrophic_2008-1}. Works within
this framework have recognized that the properties of the community
network are different from those of the entire pool. In simulations,
the mean interaction strengths were found to be smaller in the community
\citep{kokkoris_patterns_1999-1}; and certain combinations of productivities
and interspecies interactions were found appear more commonly than
others \citep{tokita_statistical_2006,yoshino_rank_2008}. No systematic
account of such differences has been provided.

\begin{figure}
\begin{centering}
\includegraphics[bb=0bp 70bp 320bp 248bp,clip,width=1\columnwidth]{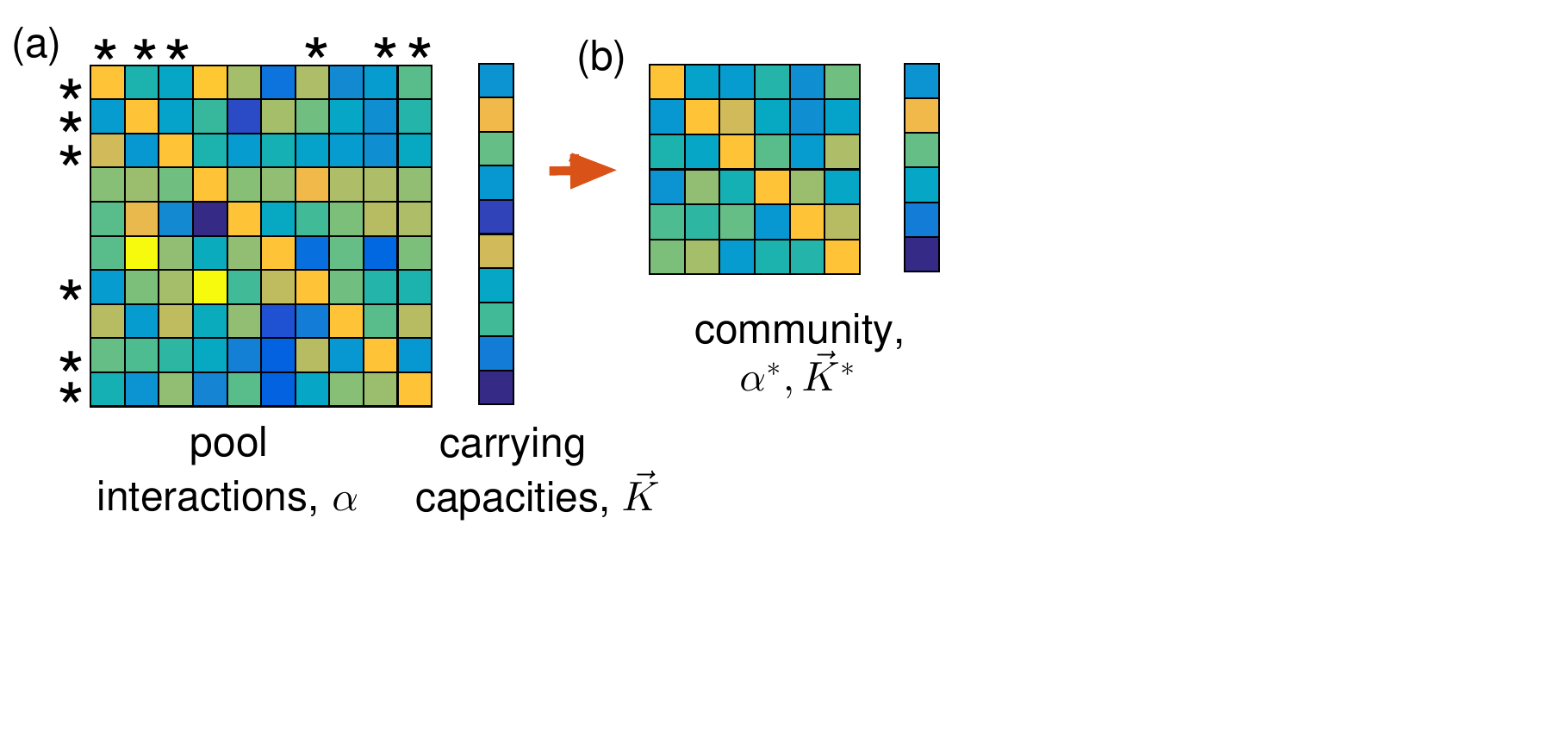}
\par\end{centering}
\caption{\label{fig:overview}The assembly process generates community networks
with properties different from in the pool. (a) The strength of interspecies
interactions between all species in the pool, and their carrying capacities.
A subset of species (marked by stars) forms the local community. (b)
The reduced matrix of community interspecies interactions, and the
reduced vector of carrying capacities. They have new statistical properties,
including changes in the elements' distributions, correlations between
interactions, and between interactions and carrying capacities.}
\end{figure}

In this work we study how network properties are influenced by the
constraints of community assembly from a given species pool, through
a systematic framework giving analytical predictions. Our aim is two
fold. First, to describe how the statistical properties of the network
are altered when restricted to interactions inside viable communities,
see Fig. \ref{fig:overview}(a,b). Even if interactions between all
pool species are ``maximally random'', comprised of a single trophic
level with random uncorrelated interactions between species, the community
networks are found to be structured. This is significant in light
of the large body of work, following \citep{may_will_1972}, that
models communities using random interactions. The changes in statistical
properties include correlations between interspecies interactions,
and between them and the carrying capacities. In addition, carrying
capacities are on average higher then in the pool, and interspecies
interactions are less competitive or more beneficial.

Secondly, focusing on the assembled community we ask: how does its
network allow all community species to coexist? After all, the persistence
of the community species is somehow encoded in the network structure.
It turns out that correlations between interactions play a crucial
role in allowing for higher diversity. But finally, it is a simple
relation between interspecies interactions and species abundances
that fully accounts for the coexistence of all species in the community.

The model includes a pool of species, a subset of which forms the
community, whose members are the persistent species (those whose abundance
does not decay to zero). The community must be resistant to invasion
by pool species outside it, so that an invader's abundance decays
if it is introduced in small abundance. This criterion accounts for
the effect of migration, if migration acts on long enough time-scales
which allow the community to relax between colonization attempts \citep{drake_mechanics_1990,case_invasion_1990,law_permanence_1996,bastolla_biodiversity_2005,fried_communities_2016}.
The dynamics of the abundances $N_{i}$ of the $S$ species in the
pool, with $i=1..S$, are modeled by the generalized Lotka-Volterra
equation
\begin{equation}
\frac{dN_{i}}{dt}=\frac{r_{i}}{K_{i}}N_{i}\left(K_{i}-N_{i}-\sum_{j,(j\neq i)}\alpha_{ij}N_{j}\right)\ .\label{eq:LV}
\end{equation}
$r_{i}$ are the intrinsic growth-rates, $K_{i}$ are the carrying
capacities, and $\alpha_{ij}$ for $i\neq j$ encode interspecies
interactions with positive values representing competition. The analytical
techniques can be applied to a broad class of other models.

In order to proceed, the parameters $r_{i},K_{i}$ and $\alpha_{ij}$
of the species in the pool \textendash{} as would be measured in the
local conditions \textendash{} need to be specified. Assuming that
detailed information on all these parameters is not available, we
turn to a null model in which they are sampled at random. Ever since
the pioneering work of May \citep{may_will_1972}, models with random
parameters have played an important role in theoretical ecology. However,
in contrast to Ref. \citep{may_will_1972}, here the community interactions
are not drawn at random, and characterizing their emergent structure
is the aim of this work. Using a simple null model for the pool allows
to disentangle the effects of the assembly process from other factors
influencing interaction patterns, for example the mechanisms that
generate the interactions such as competition of resources.

The interspecies interactions are thus sampled independently except
for possibly a correlation between $\alpha_{ij}$ and $\alpha_{ji}$,
set by the coefficient $\gamma\equiv\corr(\alpha_{ij},\alpha_{ji})$,
so that $-1\leq\gamma\leq1$. It is not restricted to the symmetric
case, $\gamma=1$. $\alpha_{ij}$ can have any distribution, as long
as it is not long-tailed (see Appendix \ref{appendix:derivations}
for the technical condition). Carrying capacities may vary between
species, in which case they are sampled independently. Results are
presented for a normal distribution of $K_{i}$ for which the analytical
expressions are compact; The main conclusions remain unchanged for
other distributions. By rescaling $N_{i}\rightarrow N_{i}/\mean\left(K_{i}\right)$
we set $\mean\left(K_{i}\right)=1$.

The analytical framework follows a long tradition in ecology, of adding
a species and asking whether it can invade \citep{macarthur_limiting_1967},
but goes beyond this to analyze the probability for it to invade and
its abundance if it succeeded. This is known as the cavity method
in the physics literature \citep{mezard_sk_1986,crisanti_sphericalp-spin_1993,diederich_replicators_1989,rieger_solvable_1989,opper_phase_1992},
which in ecological contexts has been used to calculate species abundance
distributions and other quantities, such as whether multiple equilibria
exist \citep{rieger_solvable_1989,opper_phase_1992,tokita_species_2004,yoshino_rank_2008,biscari_replica_1995,diederich_replicators_1989}.
Here, in order to study the community network, we note that its properties
can be obtained by conditioning on the persistence of the species
involved. For example, the probability distribution of a community
interaction strength $\alpha_{ij}^{\ast}$ is by definition given
by $\Pr\left(\alpha_{ij}^{\ast}\right)=\Pr\left(\alpha_{ij}|N_{i}>0,N_{j}>0\right)$,
where $N_{i},N_{j}$ are the abundances after a long time. This quantity
is calculated by introducing two species and asking that they both
persist. Joint distributions of multiple interactions and carrying
capacities are similarly obtained by conditioning on the persistence
of all the species involved. This provides systematic access to all
moments and marginal probability distributions of the community network.

The analytical technique is controlled at large pool sizes $S$ with
individually weak interspecies interactions, i.e. keeping the asymmetry
$\gamma$ and the parameters $\mu\equiv S\mean\left(\alpha_{ij}\right)$
and $\sigma^{2}\equiv S\var\left(\alpha_{ij}\right)$ constant. Good
agreement with numerical simulations is found for modest number of
species, see Fig. \ref{fig:alpha_mean_and_distrib},\ref{fig:two_alph_corr}
for $S=15$ and communities down to 6-7 species. Qualitative agreement
\textendash{} in particular, the sign of correlations \textendash{}
is found for all systems sizes, both for Gaussian and uniform distributions
of $\alpha_{ij}$, see Appendix \ref{appendix:small_S}. This encompasses
systems with purely competitive interactions ($\alpha_{ij}$ are all
positive), as well as ones that include mixtures of competitive and
beneficial interactions.

\section{Results}

The results section is organized as follows. First, properties of
the interspecies interactions and carrying capacities are presented.
Then the relation between network properties and species coexistence
is addressed. The results are compared with numerical simulations,
described in Appendix \ref{appendix:numerics}. Derivations are given
in Appendix \ref{appendix:derivations}. Analytical expressions are
quoted to lowest order in $1/S$.

Before turning to the community network, we briefly describe the dynamical
behavior of the model. It exhibits three distinct regimes, or `phases',
depending on the model parameters ($\mu,\sigma,\gamma$ and the distribution
of $K_{i}$), separated by sharp boundaries when $S$ is large at
fixed $\mu,\sigma$, see Fig. \ref{fig:phase_diagram_gam_0}. Details
are given in Appendix \ref{appendix:phase_diag}. The analytical results
are exact in the unique equilibrium phase, and qualitatively correct
in the multiple attractors phase, as seen by comparing with numerics.
In a third phase the abundances grow without bound; here the description
in terms of Lotka-Volterra equations probably breaks down and this
regime will not be further discussed.


\begin{figure}[ptb]
\centering{}\includegraphics[width=0.8\columnwidth]{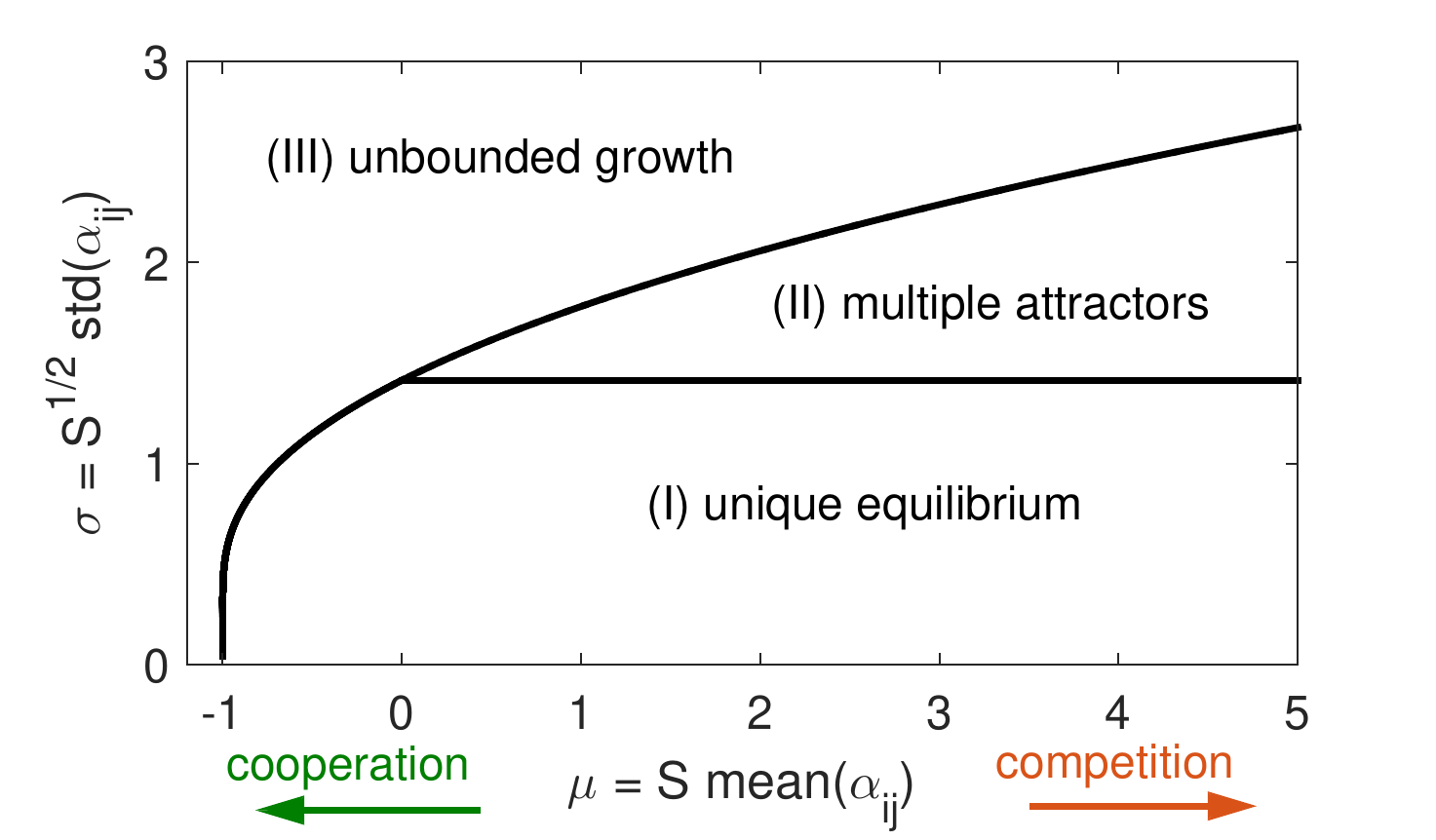}\caption{\label{fig:phase_diagram_gam_0}The model exhibits three distinct
dynamical behaviors, depending on model parameters. In phase I, a
unique stable equilibrium that is resistent to invasion exists for
any system. In phase II, multiple dynamical attractors exist, which
may be stable equilibria or other attractors such as limit cycles,
and the community composition depends assembly history. In phase III,
abundances grow without bound; Here the Lotka-Volterra equations likely
break down and this phase will not be discussed. The phases are shown
for asymmetric interactions ($\gamma=0$) and equal carrying capacities,
all set to $K_{i}=1$.}
\end{figure}


\subsection{properties of the community network}

\begin{figure}
\centering{}\includegraphics[bb=20bp 0bp 480bp 248bp,width=1\columnwidth]{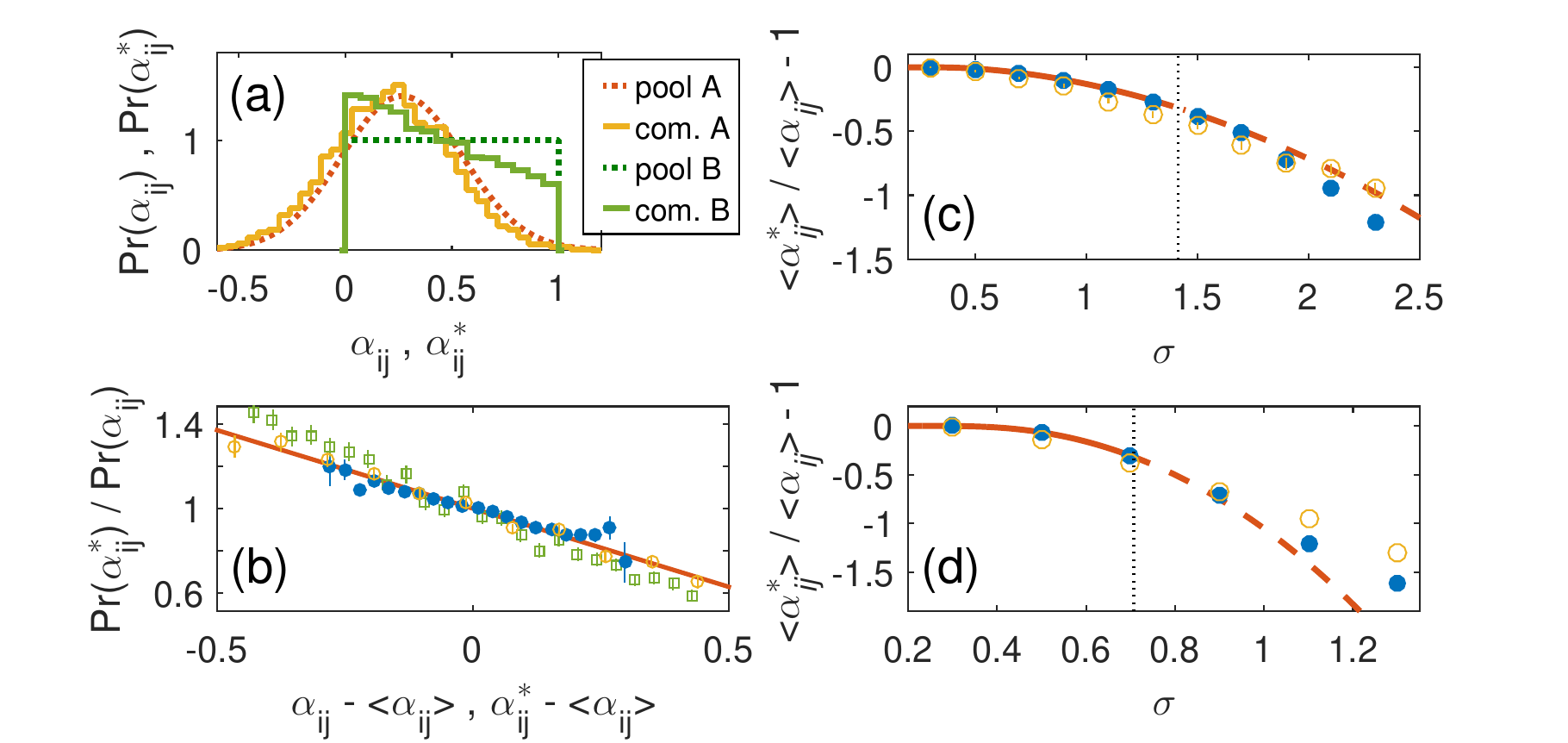}\caption{\label{fig:alpha_mean_and_distrib}Changes to interspecies interactions.
(a) The distribution of interspecies interactions (solid lines) is
enhanced at lower values \textendash{} less competitive or more beneficial
\textendash{} as compared to the pool (dotted lines). (b) The enhancement
factor is a linear function of the interaction strength. (c,d) This
shifts the mean of $\alpha_{ij}^{\ast}$, reducing mean competition,
shown for $\gamma=0$ (c) and $\gamma=1$ (d). In (b,c,d), solid and
dashed lines are analytical predictions, which fit perfectly to numerics
at large $S$ (full circles) in the unique equilibrium phase (left
of vertical dotted line in (c,d)). Open circles are numerical results
for Gaussian distribution of $\alpha_{ij}$ and $S=15$, generating
communities with down to around $6$ species. Diamonds in (b) are
uniform distribution of $\alpha_{ij}$ on $\left[0,1\right]$, and
$S=15$. Gaussian distributions run with $\mu=4$. In (b) all models
have $\sigma=\sqrt{5}/2$, as for the model with $\alpha_{ij}$ uniform
on $\left[0,1\right]$.}
\end{figure}

This section describes properties of community networks. Throughout,
$\alpha^{\ast},\vec{K}^{\ast}$ and $\vec{N}^{\ast}$ will denote
the network parameters and abundances restricted to the community
species, and $S^{\ast}$ the number of species in the community. The
pair $\alpha^{\ast},\vec{K}^{\ast}$, will have different properties
from $\alpha,\vec{K}$ of the pool. We begin with properties of $\alpha^{\ast}$.
The distribution of an element $\alpha_{ij}^{\ast}$ is a function
both of its distribution over all interactions between species in
the pool, and of the assembly process. The assembly enhances the distribution
at more negative values, corresponding to weaker competition or stronger
beneficial interactions. This enhancement is linear in interaction,
$\Pr\left(\alpha_{ij}^{\ast}=\alpha\right)/\Pr\left(\alpha_{ij}=\alpha\right)=1-c\cdot\alpha$,
see Fig. \ref{fig:alpha_mean_and_distrib}(a,b). The prefactor $c$
depends on the model parameters, and is given in Appendix \ref{appendix:derivations}.
This change shifts the overall mean, $\left\langle \alpha_{ij}^{\ast}\right\rangle $,
towards lower competition than in the pool, Fig. \ref{fig:alpha_mean_and_distrib}(b,c).
A drop in the mean competition has been described \citep{kokkoris_patterns_1999-1},
using simulations. The enhancement of weak competitive links is in
line with arguments for the prevalence and importance of weak links
\citep{paine_food-web_1992,mccann_weak_1998,kokkoris_patterns_1999-1,neutel_stability_2002}.
Note however that beneficial interactions, if they exist in the pool,
would be more enhanced when their strength $\left|\alpha_{ij}^{\ast}\right|$
is larger.

In Fig. \ref{fig:alpha_mean_and_distrib}, as everywhere else, simulation
data fits perfectly for large $S$ in the unique equilibrium phase.

\begin{figure}
\centering{}\includegraphics[width=1\columnwidth]{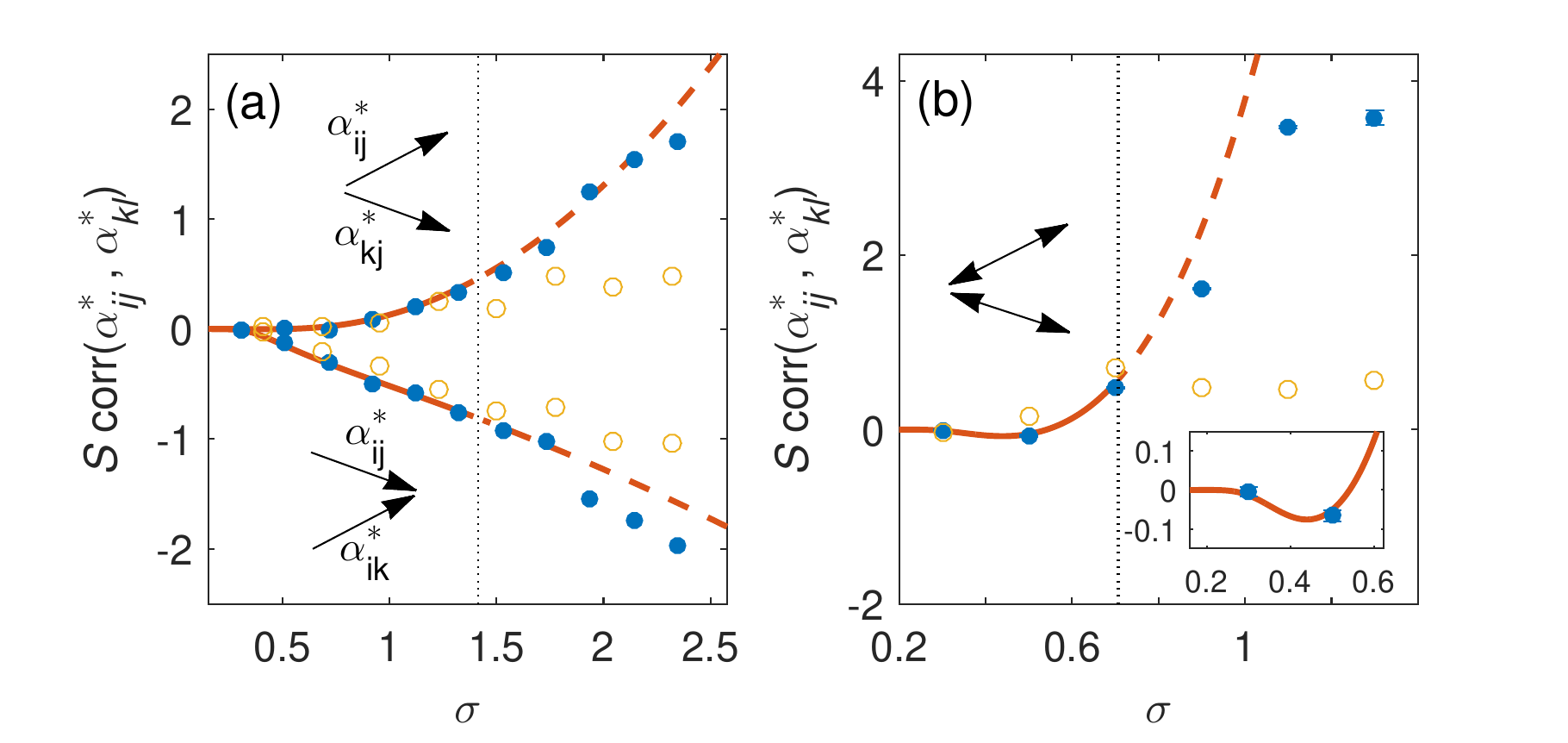}\caption{\label{fig:two_alph_corr}Correlations between interactions that share
a species, for (a) asymmetric interactions, $\gamma=0$ and (b) symmetric
interactions, $\gamma=1$. In (b) there is only one correlation between
adjacent interactions, since the interactions are symmetric. Inset
shows a part of the same graph, where the correlation changes sign.
These correlations are zero when taken over the entire species pool,
demonstrating that the community network has new structure not found
in the pool. Solid and dashed lines indicate analytical predictions.
Simulations at large $S$ (full circles) agree perfectly with theory
in phase one. Open circles are numerical results for $S=15$. Here
$\mu=4$.}
\end{figure}

Pairs of community interaction elements $\alpha_{ij}^{\ast},\alpha_{kl}^{\ast}$
that share a species (i.e. belong to the same row or column of the
matrix) are correlated, see Fig. \ref{fig:two_alph_corr}. The sign
of the correlation may depend on the role of the species in common.
For example, in asymmetric systems ($\gamma=0$), one finds a positive
correlation between $\alpha_{ij}^{\ast}$ and $\alpha_{kj}^{\ast}$,
which corresponds the influence of one species ($j$) on two others,
but negative correlations in the opposite scenario where two species
influence the same species. (Species $j$ influences $i$ with strength
$\alpha_{ij}$ through the term $-\alpha_{ij}N_{j}$ in Eq. (\ref{eq:LV})
for $dN_{i}/dt$.) Symmetric interactions ($\gamma=1$) have only
one distinct type of correlation, which changes sign as a function
of model parameters, see Fig. \ref{fig:two_alph_corr}(b). Correlations
between interaction strengths that do not share a common species are
weaker (higher order in $1/S$) and are not discussed here. All these
quantities are zero when measured over $\alpha$ of the entire species
pool, demonstrating that the community interactions indeed have different
statistical properties from the pool interactions, or from any model
in which interaction strengths are sampled independently.

\begin{figure}[ptb]
\centering{}\includegraphics[width=1\columnwidth]{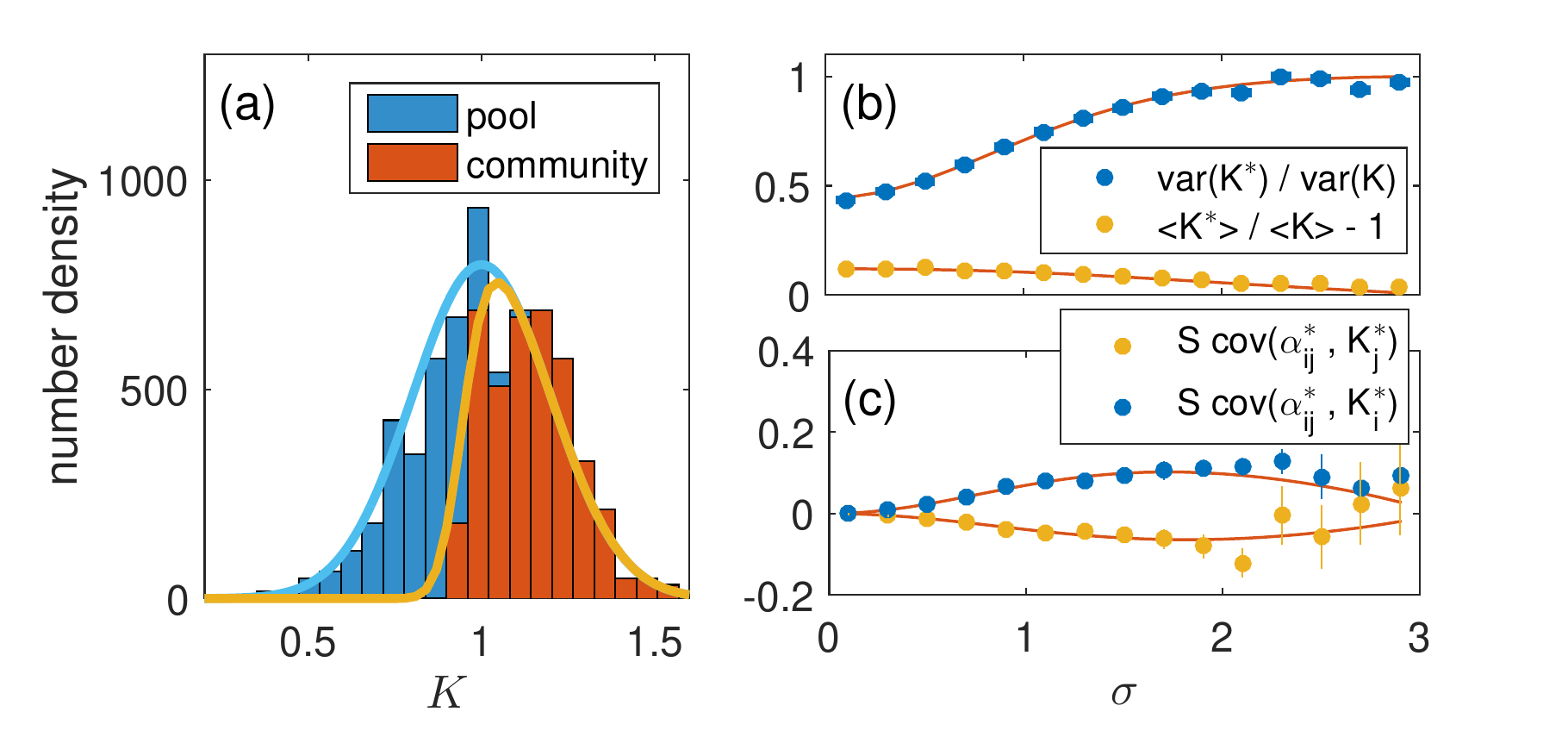}\caption{\label{fig:carrying_capacity}Change in properties of the carrying
capacities. (a) Histogram of carrying capacities in the pool and community
of a single system, compared to analytical predictions. Higher carrying
capacities are more likely to be included in the community. (b) The
community has higher mean and smaller variance of carrying capacities
than the species pool. (c) Correlations with interspecies interactions
emerge, which may be positive or negative. Solid lines are analytical
predictions. Model parameters used: $\gamma=0,\sigma_{k}=0.2$, and
in panel (a), $\sigma=0.3$.}
\end{figure}

Moving on to the carrying capacities, when the carrying capacity $K_{i}$
varies from one species to another, its distribution in the community
is altered as compared to that in the pool. This is because species
with higher carrying capacities are more likely to be included in
the community, see Fig. \ref{fig:carrying_capacity}(a). In the limiting
case of identical interspecies interactions, the persistent species
are simply those whose carrying capacity lies above some threshold.
In the other extreme, of large variability of interactions strengths
(high $\sigma$), the carrying capacities have a negligible effect
on which species persist, and so their variance is unchanged. This
`filtering' increases the mean of the carrying capacities with respect
to the pool, see Fig. \ref{fig:carrying_capacity}(b). The variance
of the distribution may change in either direction. For a Gaussian
distribution it is always reduced, see Fig. \ref{fig:carrying_capacity}(b);
This is expected to happen in similarly-shaped distributions. In other
cases the variance may increase, see Appendix \ref{appendix:derivations}
for an example. Smaller variance allows for greater maximal species
diversity \citep{bastolla_biodiversity_2005}; It is interesting that
the community assembly can act to either reduce or increase the variance.
Correlations between interspecies interactions and carrying capacities
emerge in the community. Their sign depends on the model parameters
and whether the carrying capacity of the influencing or influenced
species is included, see Fig. \ref{fig:carrying_capacity}(c), (and
also Fig. \ref{fig:carrying_capacity_gam_1} in Appendix \ref{appendix:derivations}
for $\gamma=1$).

When different species have different carrying capacities, properties
of that distribution are altered by the community assembly process.
Feasibility restricts the possible combinations of $\alpha^{\ast}$
and $\vec{K}^{\ast}$, since in equilibrium $\alpha^{\ast}\vec{N}^{\ast}=\vec{K}^{\ast}$
(with $\alpha_{ii}^{\ast}=1$) and community abundances must be positive.
This set of conditions is at the basis of many theoretical arguments
\citep{bastolla_biodiversity_2005,rohr_structural_2014,bastolla_architecture_2009}.
However, for many purposes it is desirable to have more detailed relations
between $\alpha^{\ast}$ and $\vec{K}^{\ast}$ than this set of inequalities.
The community assembly model used here has the advantage of producing
explicit predictions for the distribution of carrying capacities and
its correlations with elements of $\alpha^{\ast}$, as shown in Fig.
\ref{fig:carrying_capacity}.

\subsection{Network structure and species coexistence}

\begin{figure}
\centering{}\includegraphics[bb=0bp 0bp 490bp 248bp,clip,width=1\columnwidth]{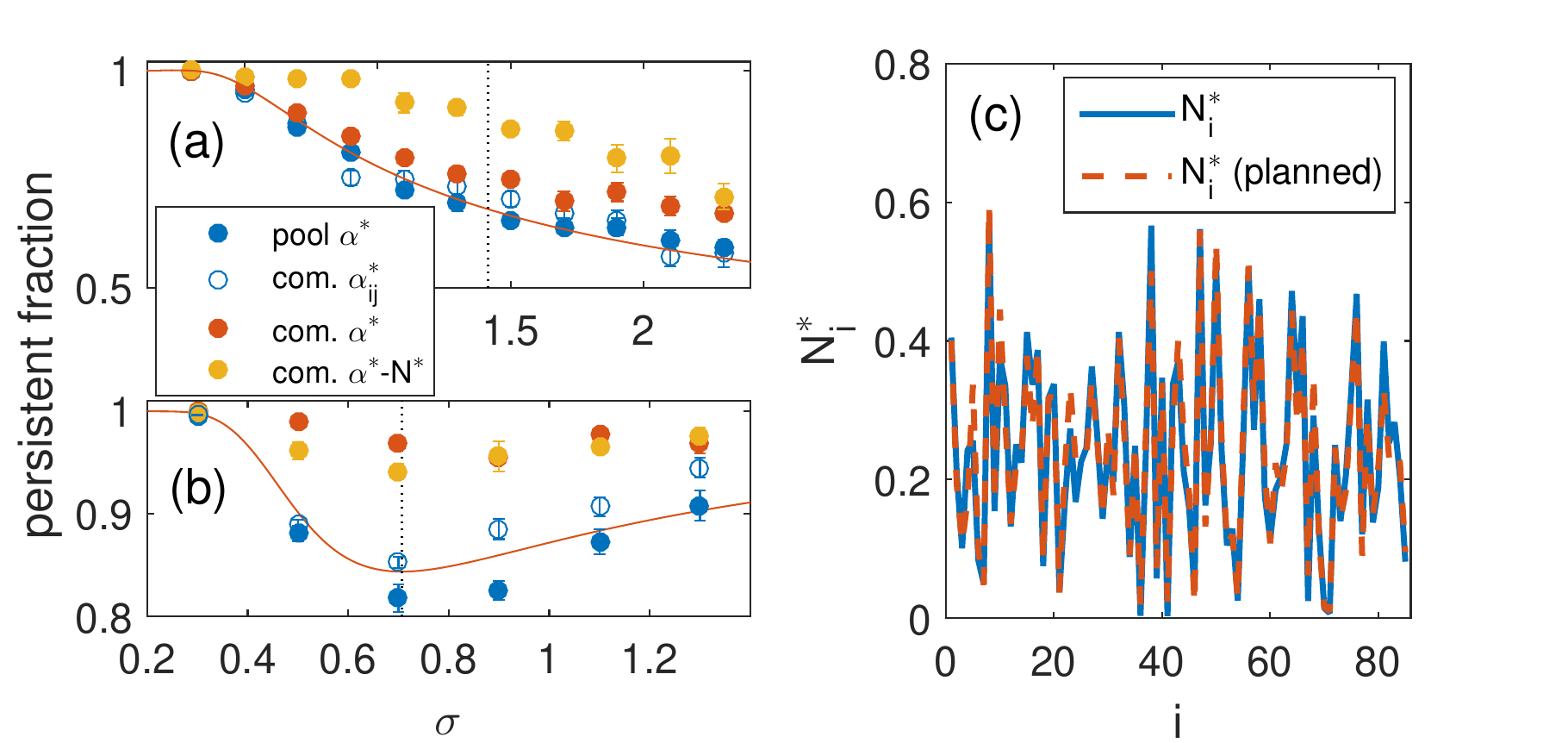}\caption{\label{fig:structure and persistence}How different properties of
the assembled network affect the number of species that persist, compared
to an assembled community. (a,b) The fraction of persistent species
in networks with different properties (for $\gamma=0,1$ respectively).
When network parameters are sampled at random, as in the pool, only
a fraction of the species persist. Sampling $\alpha_{ij}^{\ast}$
as in an assembled community does little to increase the persistence
fraction. Adding correlations between interaction elements, as in
Fig. \ref{fig:two_alph_corr}, does increase the fraction in some
cases. Consistently high diversity is obtained when correlations between
species abundances and interactions are included. (c) These correlations
are sufficient to reconstruct a given species abundance patterns.
Model parameters: $\mu=4$ with normally distributed $\alpha_{ij}^{\ast}$.
In (c), $\gamma=0$ and $\sigma=0.7$.}
\end{figure}

So far, the differences between assembled networks and those formed
by arbitrary collections of species were described. By definition,
all of the community species coexist in the community. We now ask
what properties of the assembled networks, specifically the interspecies
interactions, are responsible for this coexistence. To simplify the
presentation we focus on communities with identical carrying capacities
($K_{i}=1$), so that the community network is specified by the interactions
$\alpha^{\ast}$. To better understand the effect of network properties
on diversity (number of persistent species), we generate matrices
that are the same size as $\alpha^{\ast}$ but with different properties.
If $\alpha^{\ast}$ is replaced by completely random interactions
sampled as in the pool, only a fraction of the species persist, see
Fig. \ref{fig:structure and persistence}(a,b). Modifying the distribution
of individual interactions to match that of the assembled community
does little if anything to increase diversity. In fact it can be shown
to have no effect on large communities\footnote{As was discussed above, at large $S$ the distribution of $\alpha_{ij}^{\ast}$
has a different mean but the same variance as $\alpha_{ij}$. To leading
order in $S$ the fraction of of persist species depends only on $\sigma$
(when all $K_{i}=1$), and changes in the distribution that alter
$\mu\rightarrow\mu_{eff}$ have a sub-leading effect on the fraction
of persistent species.}. Next, including correlations between species increases the diversity,
to a degree that depends on the model parameters. To go beyond these
results and find a sufficient condition for the community species
to persist, we turn to the properties of the community interactions
$\alpha^{\ast}$ at a given species abundance $\vec{N}^{\ast}$.

\begin{figure}
\centering{}\includegraphics[bb=1050bp 150bp 4100bp 1800bp,clip,width=1\columnwidth]{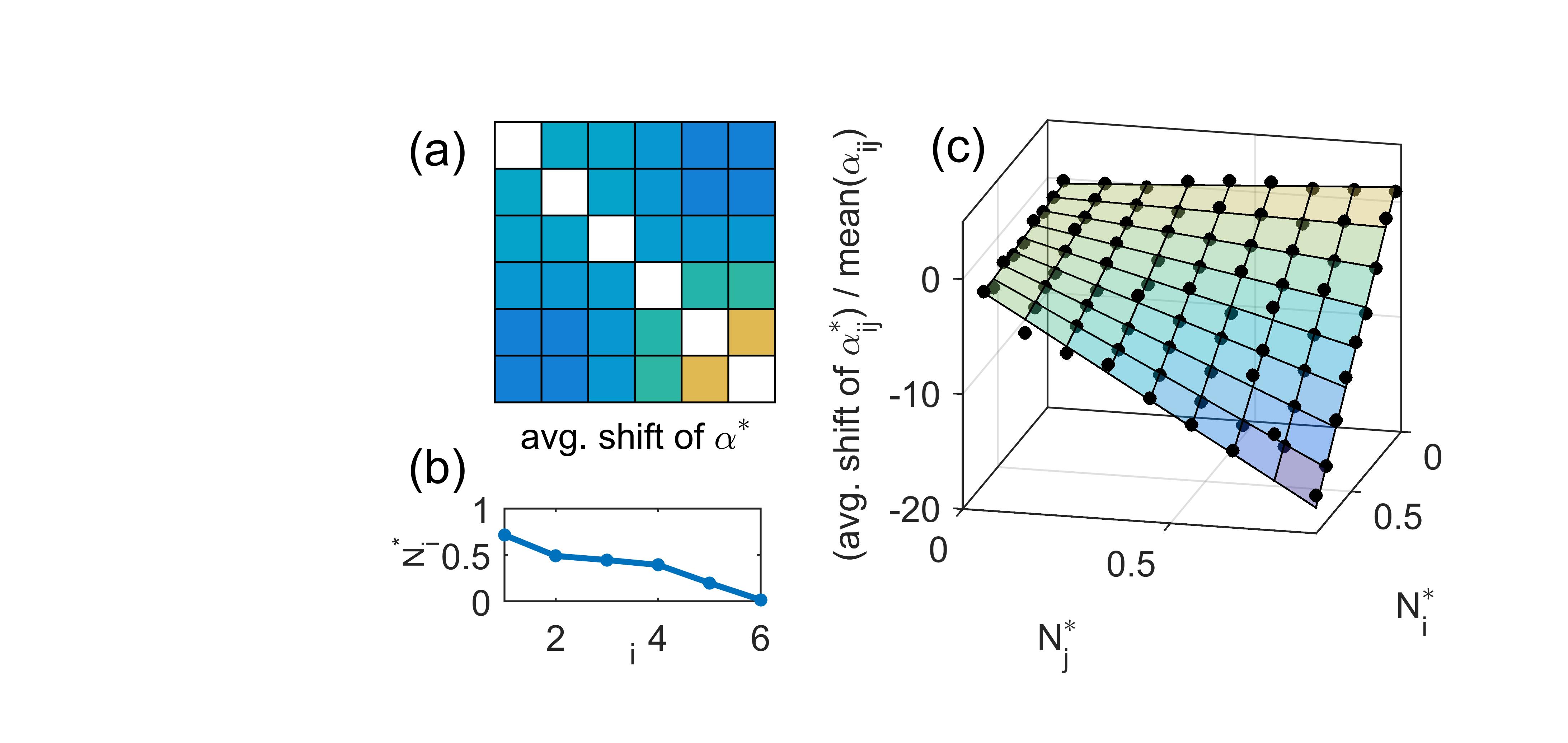}\caption{\label{fig:mean_shift}Interspecies interactions at a given species
abundance are on average shifted, as shown in (a). The shift depends
on the abundances (b), which are sorted for clarity. The bias is mostly,
but not always, towards reduced competition. This pattern is responsible
for species coexistence in the community. (c) Comparison of Eq. (\ref{eq:mean_shift})
with simulations. Model parameters: $\gamma=0$, $\sigma=1.1$ and
$\mu=4$.}
\end{figure}

As was discussed above, interspecies interactions in the community
are on average less competitive than those in the entire pool. When
considered jointly with the species abundance\footnote{In phase one (where the analytical theory is exact), the dynamics
converge to stable equilibria, so that $\vec{N}^{\ast}$ is a well-defined,
time-independent quantity.}, this change in interaction strength is not uniform, but depends
the abundance of the species involved in the interaction. Specifically,
the conditional distribution $\Pr\left(\alpha_{ij}^{\ast}|\vec{N}^{\ast}\right)$
has the same standard deviation as $\alpha_{ij}$ of the pool. Its
mean, $\mean_{\vec{N}^{\ast}}\left(\alpha_{ij}^{\ast}\right)$, is
shifted with respect to the pool mean, and depends on $N_{i}^{\ast}$
and $N_{j}^{\ast}$ in a remarkably simple way:
\begin{equation}
\frac{\mean_{\vec{N}^{\ast}}\left(\alpha_{ij}^{\ast}\right)}{\mean\left(\alpha_{ij}\right)}-1=-AN_{i}^{\ast}N_{j}^{\ast}+B\left(\gamma N_{i}^{\ast}+N_{j}^{\ast}\right)\ ,\label{eq:mean_shift}
\end{equation}
with $A=\left(1/\mu+\gamma B\left\langle N\right\rangle \right)/\left\langle N^{2}\right\rangle $
and $B=\left(1/\mu-\left\langle N\right\rangle \right)/\left\langle N^{2}\right\rangle $,
see Fig. \ref{fig:mean_shift}. The correlation coefficient of $\alpha_{ij}^{\ast},\alpha_{ik}^{\ast}$
conditioned on a given $\vec{N}^{\ast}$ is given by $\corr_{\vec{N}^{\ast}}\left(\alpha_{ij}^{\ast},\alpha_{ik}^{\ast}\right)=-\frac{1}{S\left\langle N^{2}\right\rangle }N_{j}^{\ast}N_{k}^{\ast}$.
The other distinct correlations, $\corr_{\vec{N}^{\ast}}\left(\alpha_{ij}^{\ast},\alpha_{ki}^{\ast}\right)$
and $\corr_{\vec{N}^{\ast}}\left(\alpha_{ji}^{\ast},\alpha_{ki}^{\ast}\right)$
are given by the same expression, only multiplied by $\gamma$ and
$\gamma^{2}$ respectively. From Eq. (\ref{eq:mean_shift}) one finds
that competition is always reduced when both abundances are large
and $N_{i}^{\ast}>N_{j}^{\ast}$. Depending on model parameters, competition
may increase for small $N_{i}^{\ast}$ or $N_{j}^{\ast}$; this is
visible in Fig. \ref{fig:mean_shift}(c). Expressions for a unequal
carrying capacities are very similar, see Appendix \ref{appendix:derivations}.
We note in passing that following from these results, correlations
between an interaction and the abundances of the species involved
are always negative. 

With these in mind, we generate communities by first sampling the
species abundances (whose distribution is known exactly) and then
sampling the matrix $\alpha^{\ast}$ at a given species abundance
obeying Eq. (\ref{eq:mean_shift}) and the correlations following
it. This produces communities where almost all species persist, see
Fig. \ref{fig:structure and persistence}(b,c). For large communities,
this fraction will now be shown to go to one (at least in the unique
equilibrium phase, where the theory is exact). The interactions $\alpha^{\ast}$
sampled in this way also satisfy all the properties that were described
in the first part of the Results, and in fact can be derived from
them. These relations therefore combine all community properties while
maintaining almost complete diversity.

To understand why the species persist when sampled this way, we show
that once the abundances $\vec{N}^{\ast}$ have been chosen and $\alpha^{\ast}$
sampled on their basis, the dynamics will have a fixed point at $\vec{N}^{\ast}$,
see the example in Fig. \ref{fig:structure and persistence}(c). To
show this, note that for $\vec{N}^{\ast}$ to be an equilibrium of
Eq. (\ref{eq:LV}), the quantity $I=1-N_{i}^{\ast}-\sum_{j,(j\neq i)}\alpha_{ij}^{\ast}N_{j}^{\ast}$
must vanish for all the persistent variables. Indeed, both the mean
and variance of $I$ at a given $\vec{N}^{\ast}$ are zero: in the
expectation value of $I$ , $\alpha_{ij}^{\ast}$ is replaced by $\mean_{\vec{N}^{\ast}}\left(\alpha_{ij}^{\ast}\right)$.
Using Eq. (\ref{eq:mean_shift}) one obtains\footnote{Here population averages are replaced by moments, e.g. $\left\langle N\right\rangle =S^{-1}\sum_{i}N_{i}$,
justified since correlations between abundances $N_{i}^{\ast}$ are
negligible at large $S$, see Appendix \ref{appendix:derivations}.
In addition terms of order $1/S$ have been dropped, as the expressions
for $A,B$ do not retain this level of accuracy.}
\begin{equation}
1-\mu\left\langle N\right\rangle -\mu B\left\langle N^{2}\right\rangle +\left(\mu A\left\langle N^{2}\right\rangle -\mu B\gamma\left\langle N\right\rangle -1\right)N_{i}^{\ast}
\end{equation}
which is zero, using the definitions of $A,B$. A similar calculation
using $\corr_{\vec{N}^{\ast}}\left(\alpha_{ij}^{\ast},\alpha_{ik}^{\ast}\right)$
given after Eq. (\ref{eq:mean_shift}) shows that the variance of
$I$ is also zero. And since the sampled abundances $\vec{N}^{\ast}$
are chosen to be positive, the network admits a feasible solution.
It is also stable, at least were the theory is exact, see the Discussion
section below. This completes the argument for species persistence.

Referring to the species abundances when discussing the network structure
might seem redundant, since the abundances are given by solving $\alpha^{\ast}\vec{N}^{\ast}=\vec{K}^{\ast}$
(with $\alpha_{ii}^{\ast}=1$). But the joint distribution of $\alpha^{\ast},\vec{N}^{\ast}$
would translate to moments of all orders $\alpha^{\ast}$, if written
directly in terms of $\alpha^{\ast}$ and $\vec{K}^{\ast}$. Diversity
depends on these more complicated correlations, involving the inverse
of the matrix $\alpha^{\ast}$.

The bi-linear dependence of $\mean_{\vec{N}^{\ast}}\left(\alpha_{ij}^{\ast}\right)$
on $N_{i}^{\ast}$ and $N_{j}^{\ast}$ resembles the Hebbian learning
rule for Hopfield neural networks, where a pattern to be memorized
is a vector of binary variables $\vec{\xi}$. It is memorized by adding
to the interaction strength a term proportional to $\xi_{i}\xi_{j}$.
This similarity\footnote{The closest analogy is to the symmetric Lotka-Volterra model, $\gamma=1$.
The dynamics in both models admit a Lyapunov function: $\frac{1}{2}\sum_{i}N_{i}\left(-\sum_{j}\alpha_{ij}^{\ast}N_{j}+2\right)$
for the Lotka-Volterra model and $\frac{1}{2}\sum_{i,j}w_{ij}s_{i}s_{j}$
for the Hopfield model. And the patterns $\alpha_{ij}^{\ast}=\mean_{\vec{N}^{\ast}}\left(\alpha_{ij}^{\ast}\right)$
and $w_{ij}\propto\xi_{i}\xi_{j}$ generate maxima at $\vec{N}^{\ast}$
and $\vec{\xi}$ respectively of the corresponding Lyapunov functions.} is intriguing in light of the very different mechanisms shaping the
interaction strengths: in neural networks the strength of the connections
is changed in the learning process. In contrast, the reduced $\alpha^{\ast}$
matrix is formed by keeping only the persistent species, rather than
by modifying specific matrix entries. The interpretations are also
different, as species abundance is viewed as a consequence of the
assembly process, rather than an external input to be memorized.

\section{Discussion}

\emph{What generates these properties} \textendash{} How do these
patterns emerge from the community assembly process? It is quite intuitive
that competition is on average reduced (Fig. \ref{fig:alpha_mean_and_distrib}):
Species that suffer from less competition are more likely to persist,
along with the interactions that involve these species. To estimate
the strength of this effect, note that different values of $\alpha_{ij}$
change the probability that species $i$ persists by an order of $\alpha_{ij}$
(more precisely by $\alpha_{ij}N_{j}\rho$, where $\rho\text{ is the probability density of \ensuremath{N_{i}} at \ensuremath{N_{i}\rightarrow0}}$).
The shift in the mean interaction is roughly the typical size of $\alpha_{ij}$
weighted by the probability shifts, giving $\alpha_{ij}^{2}$ or more
precisely $\var\left(\alpha_{ij}\right)$. The mean and shifts in
the mean are comparable even for large systems with many weak interactions,
in accordance with Eq. (\ref{eq:mean_shift}), since at large $S$
and fixed $\mu,\sigma$, this variance $\sigma^{2}/S$ is comparable
to the mean $\mu/S$. 

More elaborate arguments can help to understand the signs of the correlations
in Fig. \ref{fig:two_alph_corr}. For example, consider the positive
correlations between interactions sharing the same influencing species,
Fig. \ref{fig:two_alph_corr}(b). This is because a pair of interactions
is less likely to be found in the community when the effect of one
species on two others has opposing trends, causing one of the species
to suffer from stronger competition which reduces its probability
to persist. These arguments are in essence Bayesian: from the probability
that species persist given certain network patterns, one obtains the
probability of finding these patterns given that the involved species
persist.

\emph{Stability} \textendash{} Stability may be an important factor
affecting the structure of communities \citep{may_will_1972,montoya_ecological_2006}.
The main results of this paper follow from feasibility and resistance
to invasion, without invoking (linear) stability. Conversely, this
means that the results do not follow from requiring that a fixed-point
be stable. Linear stability requires that the matrix $M_{ij}^{\ast}=-\alpha_{ij}^{\ast}r_{i}^{\ast}N_{i}^{\ast}/K_{i}^{\ast}$
be negative definite. This is generally the case in phase one (see
Fig. \ref{fig:phase_diagram_gam_0}), if $\vec{r}^{\ast}$ is sampled
independently from $\alpha^{\ast}$ and $\vec{K}^{\ast}$. This was
tested for different distributions (including identical values, exponential,
power-law and uniform distributions). Stability may play a role the
second phase by selecting certain fixed-points over others.

\emph{Nestedness} \textendash{} The pattern of the mean of $\alpha^{\ast}$
at a given abundance $\vec{N}^{\ast}$, Eq. (\ref{eq:mean_shift}),
has the following property: when the rows columns are sorted by increasing
abundance, the strongest interactions concentrate in the upper left
corner (close to the element $\alpha_{11}^{\ast}$), as is visually
clear in Fig. \ref{fig:mean_shift}(a,c). Such a `nested' pattern
is commonly discussed in the context of bipartite ecological networks
with binary entries, such as mutualistic networks \citep{bascompte_nested_2003},
but can be used to describe any network \citep{jonhson_factors_2013}.
The element-to-element variations around the mean might make it difficult
to visually observe this pattern, compare Fig. \ref{fig:overview}(b),
and quantitative measures for nestedness should be used. The relation
between this phenomena and nestedness in other systems is an interesting
direction for future research.

The predictions of the theory could be tested against experiments,
if interaction strengths can be measured. A community assembly experiment
would be preferable, as it allows to directly compare between the
pool and the community. In systems where the interactions are generated
by a specific mechanism, the interactions in the pool might have different
statistics, and the calculations presented here could be carried out
for these scenarios. Indeed, at the core of the analytical technique
are objects (the desired quantities conditioned on persistence) which
can be evaluated in a wide range of models (such as the models in
\citep{galla_dynamics_2005,galla_random_2006,yoshino_statistical_2007,yoshino_rank_2008}),
including explicit resource competition, sparse or otherwise distributed
interactions, and interactions involving three or more species.

\smallskip{}

It is a pleasure to thank J. Friedman, J. Gore, M. Kardar, D. Kessler,
P. Mehta, D. Rothman and M. Tikhonov for valuable discussions. The
support of the Pappalardo Fellowship in Physics is gratefully acknowledged.

\appendix

\section{Derivations\label{appendix:derivations}}

In this appendix all derivations of the results are given. First,
the problem is set up and notation is defined. In the following section
the species abundance is calculated, along with quantities that are
used to study the network. This is followed by the distribution of
carrying capacities, the distribution and correlations of community
interspecies interactions, and correlations between carrying capacities
and interactions.

\subsection{Notation and model definition\label{appendix:definitions}}

Throughout, for random variable $f,g$, $P\left(f\right)$ is the
probability distribution of $f$, $\left\langle f\right\rangle $
the mean, and $\var\left(f\right)\equiv\left\langle f^{2}\right\rangle -\left\langle f\right\rangle ^{2}$,
$\cov\left(f,g\right)\equiv\left\langle fg\right\rangle -\left\langle f\right\rangle \left\langle g\right\rangle $,
$\corr\left(f,g\right)\equiv\cov\left(f,g\right)/\sqrt{\var\left(f\right)\var\left(g\right)}$
the variance, covariance and correlation coefficients. The Gaussian
distribution will be denoted by 
\begin{equation}
g\left(x;\mu,\sigma^{2}\right)\equiv\frac{1}{\sqrt{2\pi}\sigma}e^{-\frac{1}{2\sigma^{2}}\left(x-\mu\right)^{2}}\ .\label{eq:Gauss_func_def}
\end{equation}
Similar notation, $g\left(\mathbf{x};\mathbf{\mu},\Sigma\right)$,
applies to multivariate distributions where $\mathbf{\mu,x}$ are
vectors and $\Sigma$ the covariance matrix.

The Lotka-Volterra equations with varying carrying capacities, Eq.
(\ref{eq:LV}), read
\[
\frac{dN_{i}}{dt}=\frac{r_{i}}{K_{i}}N_{i}\left(K_{i}-N_{i}-\sum_{j,\left(j\neq i\right)}\alpha_{ij}N_{j}\right)\ ,
\]
It will be convenient to work with variables $a_{ij}$ defined by
\begin{equation}
\alpha_{ij}=\frac{\mu}{S}+\sigma a_{ij}\ ,\label{eq:alpha_a_def}
\end{equation}
with $\left\langle a_{ij}\right\rangle =0,\left\langle a_{ij}^{2}\right\rangle =1/S$
and $\left\langle a_{ij}a_{ji}\right\rangle =\gamma/S$, so the relations
for $\alpha_{ij}$ are satisfied. The carrying capacities $K_{i}$
are sampled independently of the $\alpha_{ij}$. By rescaling $N_{i}\rightarrow N_{i}/\left\langle K_{i}\right\rangle $,
we set $\left\langle K_{i}\right\rangle =1$. It will be convenient
to take the $K_{i}$ to be Gaussian with unit mean and $\sigma_{K}^{2}\equiv\operatorname*{var}\left(K_{i}\right)$.
Since the $K_{i}$ must be positive, this is reasonable for $\sigma_{K}\lesssim0.3$.
The calculations can be carried out for other distributions of carrying
capacities.

Fixed points $dN_{i}/dt=0$ for all $i$ require $N_{i}\left(K_{i}-N_{i}-\sum_{j\neq i}\alpha_{ij}N_{j}\right)=0$.
Using the definition of $a_{ij}$ and rearranging this becomes 
\begin{equation}
0=n_{i}\left(\lambda_{i}-un_{i}-\sum_{j\neq i}a_{ij}n_{j}+h\right)\ ,\label{eq:RE_version}
\end{equation}
where $n_{i}$ are the normalized abundances
\[
n_{i}=N_{i}/\left(\frac{1}{S}\sum_{j=1}^{S}N_{j}\right)
\]
so that $\frac{1}{S}\sum_{i=1}^{S}n_{i}=1$, and
\begin{equation}
u=\frac{1-\mu/S}{\sigma}\ \ \ \ ,\ \ \ \lambda_{i}=\frac{K_{i}-1}{\sigma\left\langle N\right\rangle }\ \ \ ,\ \ \ h=\frac{1/\left\langle N\right\rangle -\mu}{\sigma}\ .\label{eq:LV_to_RE_dictionary}
\end{equation}
From this equation it follows that $\left\langle \lambda_{i}\right\rangle =0$
and $\sigma_{\lambda}^{2}\equiv\left\langle \lambda_{i}^{2}\right\rangle =\sigma_{K}^{2}/\left(\sigma\left\langle N\right\rangle \right)^{2}$.
In this language the problem becomes: given $u$ and $\sigma_{\lambda}^{2}$,
find a value of $h$ such that: Eq. (\ref{eq:RE_version}) holds for
all species; $\sum_{i=1}^{S}n_{i}=S$; species for which $n_{i}=0$
cannot invade ($dN_{i}/dt<0$ for small $N_{i}$); and the persistent
species ($n_{i}>0$) are stable against small perturbations in $n_{i}$.
\ At large $S$, $u\simeq1/\sigma$ to lowest order in $1/S$, which
is used throughout the paper in comparisons with Lotka-Volterra simulations.
The exact form $\left(1-\mu/S\right)/\sigma$ can be used if one is
interested in the behavior close to the Hubble point \citep{kessler_generalized_2015}
$\mu/S=\mean\left(\alpha_{ij}\right)=1$.

The form used in Eq. (\ref{eq:RE_version}) has a number of advantages.
First, it is more convenient to work with zero mean and standardized
variance $a_{ij}$ variables. Secondly, a connection to the Replicator
Equation is established, as the fixed points of Eq. (\ref{eq:RE_version})
are precisely those of the Replicator Equations $dn_{i}/dt=n_{i}\left(\lambda_{i}-un_{i}-\sum_{j\neq i}a_{ij}n_{j}+h\right)$
and the connection is made to works that study its properties, such
as species abundance or the existence of multiple equilibria\footnote{The mapping in Eq. (\ref{eq:RE_version}) only holds for the equilibrium
properties. It is \emph{not} the well-known mapping of the full dynamics
\citep{hofbauer_evolutionary_1998,tokita_species_2004}, which requires
changing variables in a way that generates statistical dependencies
between interactions, even if they don't exist in the original Lotka-Volterra
equations.}. Perhaps most importantly, as Eq. (\ref{eq:RE_version}) depends
on the original parameters only through the combinations in Eq. (\ref{eq:LV_to_RE_dictionary}),
new relations are revealed. For example, if all carrying capacities
are identical ($K_{i}=1$), then $\sigma_{\lambda}^{2}=0$ and the
problem depends on all the original parameters through $u$, equal
to $1/\sigma$ at large $S$. Therefore all properties of the normalized
abundances $n_{i}$ do not depend on $\mu$, see Fig. \ref{fig:ps_n_n2_gam_0}(a,c).
This applies to all properties of the reduced matrix, whose structure
is determined by which $n_{i}$ are positive.

\emph{Summary of notation} \textendash{} The set of persistent variables,
for which $N_{i}>0$, are denoted by $N_{i}^{\ast}$, and similarly
$K_{i}^{\ast},n_{i}^{\ast},\lambda_{i}^{\ast}$. $S^{\ast}$ will
denote the size of the community (number of persistent variables).
The community (or reduced)\ interaction matrix is $\alpha^{\ast}$,
containing all interactions $\alpha_{ij}$ for which $N_{i},N_{j}>0$.
The fraction of persistent variables and the second moment will be
denoted by 
\[
\phi\equiv S^{\ast}/S\ \ \ \ ;\ \ \ \ q\equiv\left\langle n_{i}^{2}\right\rangle \ .
\]
Note that the average in $q$ also includes species for which $n_{i}=0$.
The correlations between abundances $n_{i}$ for different species
are weak (as shown below), and $S^{\ast}$ fluctuations will be of
order $\sqrt{S^{\ast}}$, so at large $S$, $\phi$ can be replaced
with its mean value. Similarly, $\sum_{i}n_{i}^{k}/S$ can be replaced
with $\left\langle n_{i}^{k}\right\rangle $.

Because $n_{i}=0$ for species outside the community, $\sum_{i^{\prime}}n_{i^{\prime}}^{\ast}=\sum_{i}n_{i}=S$,
therefore 
\[
\left\langle n_{i}^{\ast}\right\rangle =\sum_{j}n_{j}^{\ast}/S^{\ast}=1/\phi\ ,
\]
and similarly 
\[
\left\langle \left(n_{i}^{\ast}\right)^{2}\right\rangle =q/\phi\ .
\]

It will be useful to consider the change in a solution $n_{i}$ to
Eq. (\ref{eq:RE_version}) as the $\lambda_{i}$ are varied,
\[
v\equiv\left\langle \frac{\partial n_{i}}{\partial\lambda_{i}}\right\rangle 
\]
and use the shorthand notation
\[
\hat{u}\equiv u-\gamma v\ .
\]

\subsection{Species abundance distribution\label{appendix:SAD}}

Here a variant of the cavity method \citep{mezard_sk_1986,crisanti_sphericalp-spin_1993,diederich_replicators_1989,rieger_solvable_1989,opper_phase_1992}
is used. It is based on the dynamical cavity method \citep{rieger_solvable_1989,opper_phase_1992},
which does not require $\alpha_{ij}$ to be symmetric, but replaces
its generating functional formalism by a more elementary derivation,
close in spirit to \citep{mezard_sk_1986}. It proceeds by adding
a new species along with newly sampled interactions with the existing
system, and comparing the properties of the solution with $S$ species
to that with $S+1$ species, requiring that the new species has the
same properties as the rest.

Assume that the abundances $n_{i\backslash0}\,$ of the species in
the pool $i=1..S$ are known. Introduce a new species with interactions
$\left\{ a_{0i},a_{i0}\right\} _{i=1..S}$ and $\lambda_{0}$. For
the purposes of the derivation, Eq. (\ref{eq:RE_version}) is extended
to include an additional small perturbations $\xi_{i}$ to the $\lambda_{i}$
of each species, later set to zero:
\begin{equation}
0=n_{i}\left(\lambda_{i}-un_{i}-\sum_{j\neq i}a_{ij}n_{j}+h+\xi_{i}\right)\label{eq:RE_version_with_response}
\end{equation}
and the response to the perturbation is
\begin{equation}
v_{ij}\equiv\left[\partial n_{i}/\partial\xi_{j}\right]_{\xi_{j}=0}\ .\label{eq:v_ij_def}
\end{equation}

Once the new species is introduced, it might invade and its final
abundance will be $n_{0}>0$, or else $n_{0}=0$. The effect it has
on the species $i\geq1$ is\footnote{Here we assume that $v_{ij}=0$ if $n_{i}=0$. A small fraction, of
order $1/\sqrt{S}$, of the species with $n_{j\backslash0}$ may acquire
a positive abundance of order $1/\sqrt{S}$, but this effect is negligible.} 
\[
un_{i}=\lambda_{i}-\sum_{j\neq i}a_{ij}n_{j}+h-a_{i0}n_{0}
\]
This is the same as Eq. (\ref{eq:RE_version_with_response}), with
$\xi_{i}=-a_{i0}n_{0}$. For large $S$ each $a_{i0}n_{0}$ is small
(scales as $1/\sqrt{S}$) so that linear response can be used, $n_{j}=n_{j\backslash0}-\sum_{k}v_{jk}\xi_{k}$,
giving 
\[
n_{j}=n_{j\backslash0}-n_{0}\sum_{k}v_{jk}a_{k0}\ .
\]
If $n_{0}>0$ we substitute this equation into $0=\lambda_{0}-un_{0}-\sum_{j}a_{0j}n_{j}+h+\xi_{0}$
and rearrange to find that $n_{0}=n_{0}^{+}$, where
\begin{equation}
n_{0}^{+}\equiv\frac{\lambda_{0}-\sum_{j}a_{0j}n_{j\backslash0}+h+\xi_{0}}{u-\sum_{j,k}v_{jk}a_{0j}a_{k0}}\label{eq:n0_plus_definition}
\end{equation}
The denominator of this equation will be a finite number with negligible
fluctuations. To see this, note that $v_{ii}=O\left(S^{0}\right)$,
while $v_{ij}$ which is mediated by the $a$ interactions is expected
to be $v_{ij}=O\left(S^{-1/2}\right)$ (as can be verified later\footnote{From the definition of $v_{ij}$, Eq. (\ref{eq:v_ij_def}), the change
$\delta n_{i}$ in response to a perturbation vector $\overrightarrow{\xi}$
is $\delta n_{i}=\sum_{j}v_{ij}\xi_{j}$. If the elements of $\overrightarrow{\xi}$
are sampled independently then $\left\langle \left(\delta n\right)^{2}\right\rangle /\left\langle \xi^{2}\right\rangle =v^{2}+\left(S-1\right)\left\langle v_{ij}^{2}\right\rangle $.
As long as this is finite, as discussed in \ref{appendix:phase_diag},
$v_{ij}$ scales as $1/\sqrt{S}$.}). The sum over the $j=k$ terms in $\sum_{j,k}v_{jk}a_{0j}a_{k0}$
gives 
\[
\left\langle \sum_{j}v_{jj}a_{0j}a_{j0}\right\rangle =\sum_{j}v_{jj}\left\langle a_{0j}a_{j0}\right\rangle =\frac{\gamma}{S}\sum_{j}v_{jj}
\]
with $O\left(S^{-1/2}\right)$ fluctuations, while the sum over the
$j\neq k$ terms is $O\left(S^{-1/2}\right)$. Together, up to $O\left(S^{-1/2}\right)$
fluctuations, the denominator is equal to $u-\gamma v$ with $v\equiv\left\langle v_{jj}\right\rangle $.
All in all, the feedback of the existing species on the new species
changes the denominator from $u$ to $u-\gamma v$.

Turning to the numerator of Eq. (\ref{eq:n0_plus_definition}), the
term $\lambda_{0}-\sum_{j}a_{0j}n_{j\backslash0}+h$ has mean $h$
and variance $\sigma_{\lambda}^{2}+\sum_{j}\left\langle a_{0j}^{2}\right\rangle \left\langle n^{2}\right\rangle =\sigma_{\lambda}^{2}+q$
where $q\equiv\left\langle n^{2}\right\rangle $. This follows from
the distributions of $\lambda_{0}$ and $a_{0j}$ (all independent
from each other by construction). As a sum of many weakly correlated
terms $-\sum_{j}a_{0j}n_{j\backslash0}$ is Gaussian (see e.g. \citep{nishimori_statistical_2001}),
and so the numerator is Gaussian, $h+\xi_{0}+\sqrt{q+\sigma_{\lambda}^{2}}z$
with $P\left(z\right)=g\left(z;0,1\right)$. Setting $\xi_{0}=0$,
Eq. (\ref{eq:n0_plus_definition}) becomes
\begin{equation}
n_{0}^{+}=\frac{1}{u-\gamma v}\left(h+\sqrt{q+\sigma_{\lambda}^{2}}z\right)\ .\label{eq:n0_of_z}
\end{equation}
From the Lotka-Volterra equations, Eq. (\ref{eq:LV}), it follows
that if $n_{0}^{+}>0$, the solution $n_{0}=0$ is not stable against
invasion ($dN_{0}/dt>0$ at $N_{0}\rightarrow0^{+}$), so $n_{0}=n_{0}^{+}$.
This is where the resistance to invasion enters. Together $n_{0}=\max\left(0,n_{0}^{+}\right)$
with $n_{0}^{+}$ given in Eq. (\ref{eq:n0_of_z}). But once species
`0' has been added to the system it is in no way different from the
other species, so we may drop the subscript `0' to obtain the species
abundance distribution of all species, 
\begin{equation}
n=\max\left(0,\frac{h+\sqrt{q+\sigma_{\lambda}^{2}}z}{u-\gamma v}\right)\ .\label{eq:SAD}
\end{equation}
The distribution of $n$ is therefore a truncated Gaussian.

It remains to find the values of $q,v,h,\phi$. Using Eq. (\ref{eq:SAD})
for $n$, the relations $1=\left\langle n\right\rangle ,q=\left\langle n^{2}\right\rangle ,\phi=\left\langle \Theta^{+}\left(n\right)\right\rangle $
can be used. $\phi=\left\langle \Theta^{+}\left(n\right)\right\rangle $
is the fraction of persistent species, and $\Theta^{+}\left(n\right)=0$
if $n<0$ and $1$ otherwise. Denoting $w_{k}\left(\Delta\right)\equiv\int_{-\Delta}^{\infty}\frac{1}{\sqrt{2\pi}}e^{-\frac{z^{2}}{2}}\left(z+\Delta\right)^{k}dz$,
these relations read
\begin{align}
v\left(u-\gamma v\right) & =w_{0}\left(\Delta\right)\nonumber \\
u-\gamma v & =\sqrt{q+\sigma_{\lambda}^{2}}w_{1}\left(\Delta\right)\nonumber \\
\left(u-\gamma v\right)^{2} & =\left(1+\sigma_{\lambda}^{2}/q\right)w_{2}\left(\Delta\right)\label{eq:order_param_eqns}
\end{align}
where $\Delta\equiv h/\sqrt{q+\sigma_{\lambda}^{2}}$. A fourth equation
is obtained by differentiating Eq. (\ref{eq:n0_plus_definition})
with respect to $\xi_{0}$: if $n_{0}>0$ it gives $v_{00}=1/\left(u-\gamma v\right)$
and otherwise $v_{00}=0$. Together
\begin{equation}
v=\left\langle v_{00}\right\rangle =\phi\frac{1}{u-\gamma v}\;.\label{eq:order_param_v_eqn}
\end{equation}
This completes the set of four coupled equations for the unknowns
$q,v,h,\phi$. Using the identity $w_{2}\left(\Delta\right)=w_{0}\left(\Delta\right)+\Delta\cdot w_{1}\left(\Delta\right)$
and the definition of $\Delta$, we also have 
\begin{equation}
h=q\left[u-v\left(1+\gamma+\sigma_{\lambda}^{2}/q\right)\right]\ .\label{eq:h_relation}
\end{equation}
 These equations were first derived, for $\sigma_{\lambda}^{2}=0$,
in the context of the Replicator Equations in \citep{diederich_replicators_1989,opper_phase_1992}.
They can be solved numerically by evaluating $q=w_{2}/w_{1}^{2},v=w_{0}/\left(w_{1}\sqrt{q+\sigma_{\lambda}^{2}}\right)$
and $u=\gamma v+\sqrt{q+\sigma_{\lambda}^{2}}w_{1}$ as functions
of $\Delta$ and $\sigma_{\lambda}^{2}$, and then plotting the different
quantities against each other.

\begin{figure}[ptb]
\centering{}\includegraphics[bb=30bp 0bp 470bp 230bp,width=1\columnwidth]{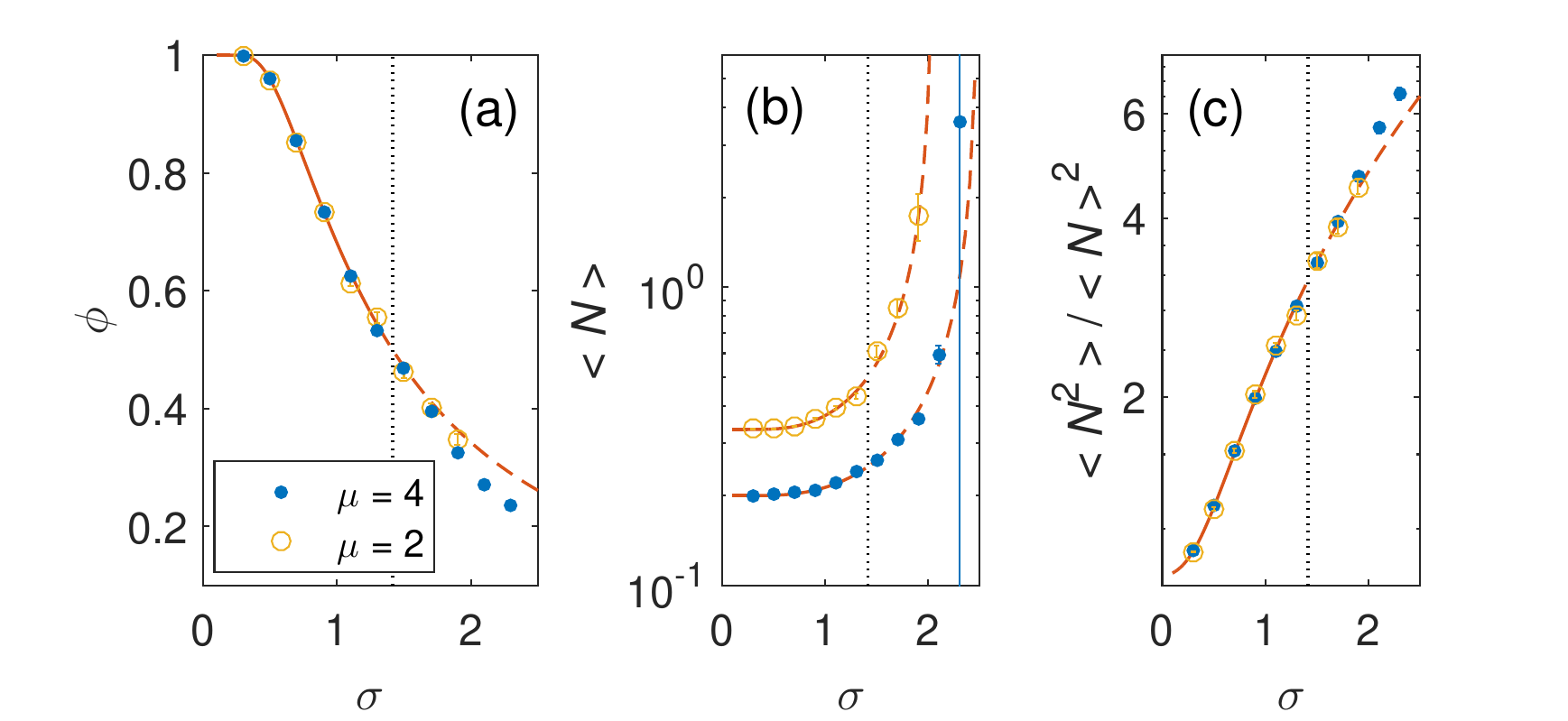}\caption{Properties of species abundance, for $\gamma=0,$ $\mu=2,4$ and $\sigma_{K}^{2}=0$.
$\phi$ is the fraction of persistent species. Solid and dashed line
are analytical predictions, exact in the unique equilibrium phase,
left of the vertical dotted line. }
\label{fig:ps_n_n2_gam_0}
\end{figure}

\begin{figure}[ptb]
\centering{}\includegraphics[width=1\columnwidth]{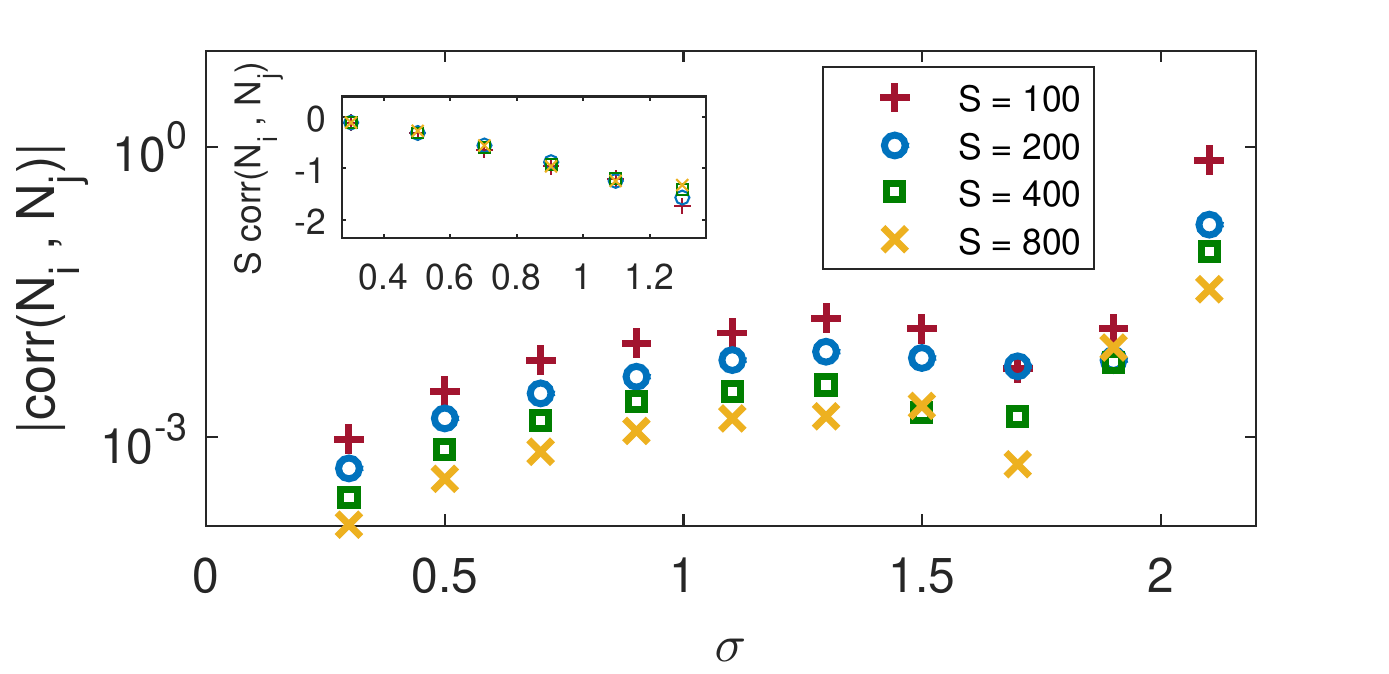}\caption{Correlation between abundances. Inset: In phase one, the correlation
for different $S$ collapse when multiplied by $S$. }
\label{fig:ni_nj_corr}
\end{figure}

Returning to the Lotka-Volterra variables $N_{i}=\left\langle N\right\rangle n_{i}$,
one has $\left\langle N\right\rangle =\sigma h+\mu$ from Eq. (\ref{eq:LV_to_RE_dictionary})
and $\left\langle N^{2}\right\rangle =q\left\langle N\right\rangle ^{2}=q\left(\sigma h+\mu\right)^{2}$.
The species abundance of $N_{i}$ is a truncated Gaussian from Eq.
(\ref{eq:SAD}), fully characterized by $\left\langle N\right\rangle $
and $\left\langle N^{2}\right\rangle $. Fig. \ref{fig:ps_n_n2_gam_0}
shows the fraction\ of persistent variables and the moments $\left\langle N\right\rangle $
and $\left\langle N^{2}\right\rangle $ and the species abundance
distribution, for $\gamma=0,\sigma_{K}^{2}=0$ and $\mu=2,4$. Those
are compared with numerical simulations at large $S$ (the simulations
are described in Sec. \ref{appendix:numerics}. Small $S$ values
are discussed in Sec. \ref{appendix:small_S}). The analytical results
are exact phase one, where a unique equilibrium exists (left of the
vertical dotted line), but serve as a good approximation beyond that.
Note also that the quantities in Fig. \ref{fig:ps_n_n2_gam_0}(a,c),
which are properties of the normalized abundances $n_{i}$ alone,
give the same result for both $\mu=2$ and $4$, as was discussed
in Sec. \ref{appendix:definitions}. The correlations between abundances
are weak and higher order in $1/S$, see Fig. \ref{fig:ni_nj_corr}.
Their precise form will not be needed in the following.

\subsection{Carrying capacity distribution of persistent community\label{appendix:carry_capac}}

As a first calculation of a property of the community network, the
distribution of carrying capacity in persistent community is derived.
It will turn out to have higher average values and lower variance
in the community as compared the entire species pool, see Fig. \ref{fig:carrying_capacity}
in the main text.

\begin{figure}[ptb]
\centering{}\includegraphics[width=0.5\columnwidth]{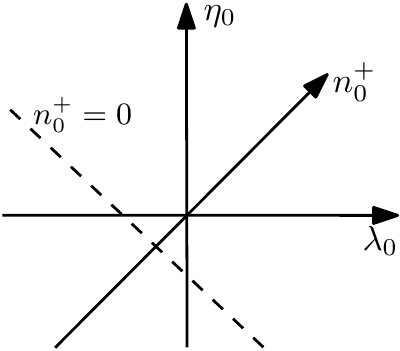}\caption{$\lambda_{0}$ is the rescaled carrying capacity of the added species
$0$, $\eta_{0}$ the accumulated effect of interspecies interactions,
and $n_{0}$ is the normalized abundance. $\lambda_{0}$ and $\eta_{0}$
are independent random variables, from which the joint distribution
of $\lambda_{0}$ and $n_{0}$ is calculated, in the region where
$n_{0}=n_{0}^{+}>0$. }
\label{fig:eta_lambda_diagram}
\end{figure}


As in the previous section, see Eq. (\ref{eq:n0_of_z}), $n_{0}=\max\left(0,n_{0}^{+}\right)$
with
\[
n_{0}^{+}=\frac{1}{\hat{u}}\left(\lambda_{0}+\eta_{0}+h\right)\ .
\]
Here $\eta_{0}\equiv-\sum_{j}a_{0j}n_{j\backslash0}$ and $\hat{u}\equiv u-\gamma v$.
Recall that $P\left(\lambda_{0}\right)=g\left(\lambda_{0};0,\sigma_{\lambda}^{2}\right)$,
$P\left(\eta_{0}\right)=g\left(\eta_{0};0,q\right)$ and $P\left(\lambda_{0},\eta_{0}\right)=P\left(\lambda_{0}\right)P\left(\eta_{0}\right)$.
We now wish to obtain the joint probability of $\lambda_{0},n_{0}^{+}$,
restricted to $n_{0}^{+}>0$, see Fig. \ref{fig:eta_lambda_diagram}.
First, changing variables $\left(\lambda_{0},\eta_{0}\right)\rightarrow\left(\lambda_{0},n_{0}^{+}\right)$,
\begin{align*}
P\left(\lambda_{0},n_{0}^{+}\right) & =\left\vert \partial\eta_{0}/\partial n_{0}^{+}\right\vert P\left(\lambda_{0},\eta_{0}\right)\\
 & =\left\vert \hat{u}\right\vert g\left(\lambda_{0};0,\sigma_{\lambda}^{2}\right)g\left(\hat{u}n_{0}^{+}-\lambda_{0};h,q\right)\ .
\end{align*}
where $g\left(..\right)$ is the normal distribution, see Eq. (\ref{eq:Gauss_func_def}).
Rearranging we find
\begin{align*}
P\left(\lambda_{0},n_{0}^{+}\right) & =\frac{\left\vert \hat{u}\right\vert }{2\pi\sigma_{\lambda}\sqrt{q}}e^{-\frac{1}{2}\left[\frac{\lambda_{0}^{2}}{\sigma_{\lambda}^{2}}+\frac{1}{q}\left(\hat{u}n_{0}^{+}-\lambda_{0}-h\right)^{2}\right]}\\
 & =g\left(\lambda_{0};\frac{\hat{u}n_{0}^{+}-h}{1+q/\sigma_{\lambda}^{2}},\frac{q}{1+q/\sigma_{\lambda}^{2}}\right)P\left(n_{0}^{+}\right)
\end{align*}
where recall from the previous section that $P\left(n_{0}^{+}\right)=g\left(n_{0}^{+};\frac{h}{\hat{u}},\frac{q+\sigma_{\lambda}^{2}}{\hat{u}^{2}}\right)$.
Since $P\left(\lambda_{0},n_{0}^{+}\right)=P\left(\lambda_{0}|n_{0}^{+}\right)P\left(n_{0}^{+}\right)$,
the first term in the second line is $P\left(\lambda_{0}|n_{0}^{+}\right)$.
Now, restricting the distribution to $n_{0}>0$, the distributions
$P_{n_{0}^{+}>0}\left(\lambda_{0},n_{0}^{+}\right),P_{n_{0}^{+}>0}\left(n_{0}^{+}\right)$
change only their normalization, and $P\left(\lambda_{0}|n_{0}^{+}\right)$
remains unchanged. Also, $n_{0}=n_{0}^{+}$ when restricted to $n_{0}>0$,
and moreover this species is not different from any other in the community,
so $P\left(\lambda_{i}^{\ast},n_{i}^{\ast}\right)=P_{n_{0}>0}\left(\lambda_{0},n_{0}^{+}\right)$
for any persistent $i$:
\begin{equation}
P\left(\lambda_{i}^{\ast}|n_{i}^{\ast}\right)=g\left(\lambda_{i}^{\ast};\frac{\hat{u}n_{i}^{\ast}-h}{1+q/\sigma_{\lambda}^{2}},\frac{q}{1+q/\sigma_{\lambda}^{2}}\right)\ .\label{eq:lam_given_n}
\end{equation}
The mean of $\lambda_{i}^{\ast}$ shifts when conditioned to $n_{i}^{\ast}$,
and the variance of $P\left(\lambda_{i}^{\ast}|n_{i}^{\ast}\right)$
is not affected by $n_{i}^{\ast}$. An advantage of the conditional
expression is that moments can be easily calculated. As $\int_{-\infty}^{\infty}d\lambda_{i}^{\ast}P\left(\lambda_{i}^{\ast}|n_{i}^{\ast}\right)\lambda_{i}^{\ast}=\frac{\hat{u}n_{i}^{\ast}-h}{1+q/\sigma_{\lambda}^{2}}$,
integrating this over $\int_{0^{+}}^{\infty}dn_{i}^{\ast}P\left(n_{i}^{\ast}\right)$
and using $\left\langle n_{i}^{\ast}\right\rangle =1/\phi$ (see Sec.
\ref{appendix:definitions}) the average reads 
\begin{equation}
\left\langle \lambda_{i}^{\ast}\right\rangle =\frac{\hat{u}/\phi-h}{1+q/\sigma_{\lambda}^{2}}\ ,\label{eq:lam_avg}
\end{equation}
and by $\int_{-\infty}^{\infty}d\lambda_{i}^{\ast}P\left(\lambda_{i}^{\ast}|n_{i}^{\ast}\right)\left(\lambda_{i}^{\ast}\right)^{2}=\left(\frac{\hat{u}n_{i}^{\ast}-h}{1+q/\sigma_{\lambda}^{2}}\right)^{2}+\frac{q}{1+q/\sigma_{\lambda}^{2}}$,
$\left\langle \left(n_{i}^{\ast}\right)^{2}\right\rangle =q/\phi$
and $\left\langle n_{i}^{\ast}\right\rangle =1/\phi$,
\begin{equation}
\var\left(\lambda_{i}^{\ast}\right)=\hat{u}^{2}\frac{q/\phi-1/\phi^{2}}{\left(1+q/\sigma_{\lambda}^{2}\right)^{2}}+\frac{q}{1+q/\sigma_{\lambda}^{2}}\ .\label{eq:lam_var}
\end{equation}

Finally, note that the distribution $P\left(\lambda_{i}^{\ast}\right)$
is precisely that of $\lambda_{0}$ when $n_{0}^{+}>0$. Integrating
Eq. (\ref{eq:lam_given_n}) over $n_{0}^{+}>0$ gives
\begin{equation}
P\left(\lambda_{i}^{\ast}\right)=g\left(\lambda_{i}^{\ast};0,\sigma_{\lambda}^{2}\right)\frac{1}{2}\left[1+\sign(\hat{u})\erf\frac{\lambda_{i}^{\ast}+h}{\sqrt{2q}}\right]\ .
\end{equation}
The expressions for the moments, Eqs. (\ref{eq:lam_avg},\ref{eq:lam_var}),
could have been obtained by integrating $P\left(\lambda_{i}^{\ast}\right)$
and relating the results to terms in Eq. (\ref{eq:order_param_eqns}).
It was more convenient to use the conditional probability since $\left\langle \left(n_{i}^{\ast}\right)^{2}\right\rangle $
and $\left\langle n_{i}^{\ast}\right\rangle $ are given directly
in terms of $\phi,q$.

The distribution of the persistent carrying capacities, $P\left(K_{i}^{\ast}\right)$
and its moments can be readily deduced from $\lambda_{i}=\frac{K_{i}-1}{\sigma\left\langle N\right\rangle }$,
see Eq. (\ref{eq:LV_to_RE_dictionary}). Thus $\left\langle K_{i}^{\ast}\right\rangle =1+\sigma\left\langle N\right\rangle \left\langle \lambda_{i}^{\ast}\right\rangle $
and $\var\left(\lambda_{i}^{\ast}\right)=\frac{1}{\sigma^{2}\left\langle N\right\rangle ^{2}}\var\left(K_{i}^{\ast}\right)$.
Fig. \ref{fig:carrying_capacity} shows $P\left(K_{i}^{\ast}\right)$
and the moments for one set of model parameters. The mean satisfies
$\left\langle K_{i}^{\ast}\right\rangle >\left\langle K_{i}\right\rangle =1$
always, since up to a normalization $P\left(\lambda_{i}^{\ast}\right)$
is equal to $P\left(\lambda_{i}\right)$ multiplied by an increasing
function. For a Gaussian distribution of $K_{i}$, $\var\left(K_{i}^{\ast}\right)<\var\left(K_{i}\right)$
(this was verified by evaluating Eq. (\ref{eq:lam_var}) over a wide
range of $\mu,\sigma_{K}$, for $-1\leq\gamma\leq1$ and $\sigma$
up to the unbounded growth phase). However this will not hold for
any distribution. For example, if $P\left(K_{i}\right)$ is bi-modal,
where most of the probability is in a low and narrow part, and a smaller
part is higher and wide. If $P\left(K_{i}^{\ast}\right)$ contains
mostly the top part, then it may have a larger variance than $P\left(K_{i}\right)$\footnote{As a proof of existence, consider a pool with 7 species: five species
with $K=0.05$, one with $K=1$ and one with $K=4$. All $\alpha_{ij}=0.2$.
Only the species with $K=1,4$ will persist, and the variance of the
carrying capacities will be larger in the community. } 

\subsection{Distribution of $\alpha_{ij}^{\ast}$\label{appendix:mean_shift}}

In this section the distribution of a single element in $\alpha_{ij}^{\ast}$
is derived. Since by definition, $P\left(\alpha_{ij}^{\ast}\right)=P\left(\alpha_{ij}|n_{i},n_{j}>0\right)$,
this conditional distribution is calculated. The derivation follows
a path similar to the previous section, but now introducing two new
species at once, denoted $i=1,2$ with abundance $n_{1,2}$. Define
\[
\tilde{h}_{k}\equiv\lambda_{k}-\sum_{j\notin\left\{ 1,2\right\} }a_{kj}n_{j\backslash\left\{ 1,2\right\} }+h\ ,
\]
It mean and variance are $\left\langle \tilde{h}_{n}\right\rangle =h$
and $\left\langle \tilde{h}_{n}^{2}\right\rangle -\left\langle \tilde{h}_{n}\right\rangle ^{2}=q+\sigma_{\lambda}^{2}$.
Following the same steps as in Sec. \ref{appendix:SAD}, one finds
that if both$\ n_{1},n_{2}>0$ then $n_{1,2}=n_{1,2}^{+}$ where 
\begin{align}
\hat{u}n_{1}^{+} & =\tilde{h}_{1}+a_{12}n_{2}^{+}\nonumber \\
\hat{u}n_{2}^{+} & =\tilde{h}_{2}+a_{21}n_{1}^{+}\label{eq:np_12_of_h}
\end{align}
Where $a_{12},a_{21}$ satisfy $\left\langle a_{12}^{2}\right\rangle =\left\langle a_{21}^{2}\right\rangle =1/S$
and $\left\langle a_{12}a_{21}\right\rangle =\gamma/S$.

Following similar steps to Sec. \ref{appendix:carry_capac}, $P\left(a_{12},a_{21},n_{1}^{+},n_{2}^{+}\right)$
is first calculated. As $P\left(a_{12},a_{21},\tilde{h}_{1},\tilde{h}_{2}\right)=P\left(a_{12},a_{21}\right)P\left(\tilde{h}_{1}\right)P\left(\tilde{h}_{2}\right)$,
\begin{equation}
P\left(a_{12},a_{21},n_{1}^{+},n_{2}^{+}\right)=JP\left(a_{12},a_{21}\right)P\left(\tilde{h}_{1}\right)P\left(\tilde{h}_{2}\right)\ ,\label{eq:Pa12a21np1np2}
\end{equation}
and $\tilde{h}_{1},\tilde{h}_{2}$ are substituted by their values
from Eq. (\ref{eq:np_12_of_h}) 
\begin{align*}
\tilde{h}_{1} & =\hat{u}n_{1}^{+}+a_{12}n_{2}^{+}\ ,\\
\tilde{h}_{2} & =\hat{u}n_{2}^{+}+a_{21}n_{1}^{+}\ .
\end{align*}
$J$ is the Jacobian of the change of variables $\left(a_{12},a_{21},\tilde{h}_{1},\tilde{h}_{2}\right)\rightarrow\left(a_{12},a_{21},n_{1}^{+},n_{2}^{+}\right)$,
\[
J=\frac{1}{\hat{u}}-\frac{1}{\hat{u}^{3}}a_{12}a_{21}\ .
\]
We now expand $P\left(\tilde{h}_{1,2}\right)$ in the parameters $a_{12},a_{21}$,
since once the moments of the equation are taken below, higher powers
of $a_{ij}$ will give higher powers in $1/S$. Expanding to first
order, $P\left(\tilde{h}_{1}\right)=g\left(\hat{u}n_{1}^{+}-a_{12}n_{2}^{+};h,\hat{q}\right)$
becomes 
\begin{align*}
P\left(\tilde{h}_{1}\right) & =g\left(\hat{u}n_{1}^{+};h,q+\sigma_{\lambda}^{2}\right)-a_{12}n_{2}^{+}g^{\prime}\left(\hat{u}n_{1}^{+};h,q+\sigma_{\lambda}^{2}\right)\\
 & =\frac{1}{\hat{u}}P\left(n_{1}^{+}\right)\left[1-\frac{a_{12}}{q+\sigma_{\lambda}^{2}}n_{2}^{+}\left(\hat{u}n_{1}^{+}-h\right)\right]
\end{align*}
where $P\left(n_{1}^{+}\right)=g\left(n_{1}^{+};h/\hat{u},\left(q+\sigma_{\lambda}^{2}\right)/\hat{u}\right)$,
see Eq. (\ref{eq:n0_of_z}).

Now $P\left(a_{12},a_{21}|n_{1}^{+},n_{2}^{+}\right)\propto P\left(a_{12},a_{21},n_{1}^{+},n_{2}^{+}\right)$,
where the proportionality includes all factors that are independent
of $a_{12,21}$. Also, if both species are included in the community,
$n_{1},n_{2}>0$, then $n_{i}^{+}=n_{i}^{\ast}$. To lowest order,
from Eq. (\ref{eq:Pa12a21np1np2})
\begin{multline}
P\left(a_{12},a_{21}|n_{1}^{\ast},n_{2}^{\ast}\right)=\\
P\left(a_{12},a_{21}\right)\left[1-\hat{u}\frac{a_{12}+a_{21}}{q+\sigma_{\lambda}^{2}}n_{1}^{\ast}n_{2}^{\ast}+h\frac{a_{12}n_{2}^{\ast}+a_{21}n_{1}^{\ast}}{q+\sigma_{\lambda}^{2}}\right]\;.\label{eq:Pa12a21_of_n}
\end{multline}
This distribution is normalized when integrated over $a_{12,21}$
since $\left\langle a_{12}\right\rangle =\left\langle a_{21}\right\rangle =0$.
Using $\left\langle a_{12}^{2}\right\rangle =1/S$ and $\left\langle a_{12}a_{21}\right\rangle =\gamma/S$,
the expectation value of $a_{12}$ reads
\begin{equation}
\mean_{n_{1,2}^{\ast}}a_{12}=-\frac{\left(1+\gamma\right)\hat{u}n_{1}^{\ast}n_{2}^{\ast}-h\left(\gamma n_{1}^{\ast}+n_{2}^{\ast}\right)}{S\left(q+\sigma_{\lambda}^{2}\right)}\ .\label{eq:mu_12}
\end{equation}
and $\mean_{n_{1,2}^{\ast}}a_{12}$ is similar, only with $1\leftrightarrow2$
indices switched. Corrections to this expression are $O\left(1/S^{2}\right)$.
The variance and correlation are unchanged by the conditioning: $\var_{n_{1,2}^{\ast}}a_{12}=1/S$,
and $\corr_{n_{1,2}^{\ast}}\left(a_{12},a_{21}\right)=\gamma$.

\begin{figure}[ptb]
\centering{}\includegraphics[width=3.5319in,height=1.7106in]{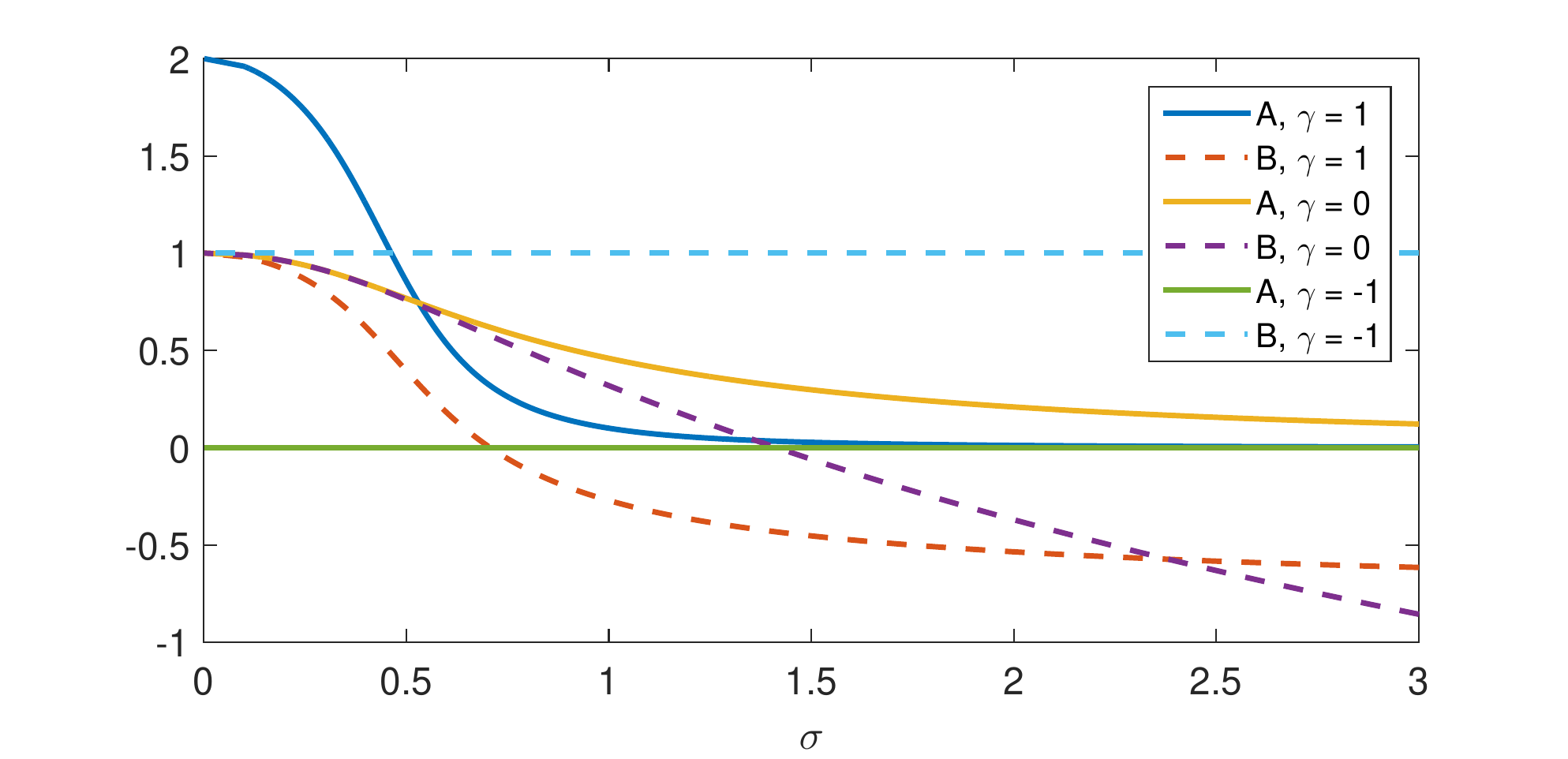}\caption{The functions $A$ and $B$ in Eq. (\ref{eq:A_B_full_expressions})
as functions of $\sigma$, for $\gamma=-1,0,1$. Here all $K_{i}=1$,
and the function $\mu A$ and $\mu B$ are plotted, since in this
case they do not depend on $\mu$.}
\label{fig:a_b}
\end{figure}

Going back to $\alpha_{ij}$, using $\alpha_{ij}=\mu/S+\sigma a_{ij}$
together with $n_{i}=N_{i}/\sum_{j=1}^{S}N_{j}$ and the definitions
of $u,h,q$ in Eqs. (\ref{eq:LV_to_RE_dictionary},\ref{eq:h_relation})
we find Eq. (\ref{eq:mean_shift}) 
\[
\frac{\mean_{\vec{N}^{\ast}}\left(\alpha_{ij}^{\ast}\right)}{\mean\left(\alpha_{ij}\right)}-1=-AN_{i}^{\ast}N_{j}^{\ast}+B\left(\gamma N_{i}^{\ast}+N_{j}^{\ast}\right)\ ,
\]
with $A,B$ given by 
\begin{align}
B & =\frac{1/\mu-\left\langle N\right\rangle }{\left\langle N^{2}\right\rangle +\sigma_{K}^{2}/\sigma^{2}}\nonumber \\
A & =\frac{\left(1+\gamma\right)\left(1/\mu+\gamma\left\langle N\right\rangle B\right)}{\left\langle N^{2}\right\rangle \left(1+\gamma\right)+\sigma_{K}^{2}/\sigma^{2}}\label{eq:A_B_full_expressions}
\end{align}
These reduce to the expressions for $A,B$ following Eq. \ref{eq:mean_shift}
in the main text when $\sigma_{K/\overline{K}}^{2}=0,\overline{K}=1$.
$A,B$ are plotted for different $\gamma$ in Fig. \ref{fig:a_b}.
Note that $A$ must vanish for $\gamma=-1$, as indeed can be seen
in the figure, since it creates is a shift of $a_{ij}$ which is symmetric
in $N_{i}^{\ast},N_{j}^{\ast}$.

The mean of $\alpha_{ij}^{\ast}$, plotted in Fig. \ref{fig:alpha_mean_and_distrib},
is 
\begin{align}
\left\langle \alpha_{ij}^{\ast}\right\rangle  & =\int_{0^{+}}^{\infty}dN_{i}^{\ast}P\left(N_{i}^{\ast}\right)\Delta\alpha_{ij}^{\ast}\nonumber \\
 & =-\frac{1}{S}\left[A\left\langle N_{i}^{\ast}N_{j}^{\ast}\right\rangle -B\left(\gamma\left\langle N_{i}^{\ast}\right\rangle +\left\langle N_{j}^{\ast}\right\rangle \right)\right]\label{eq:avg_alph}
\end{align}
and $\left\langle N_{i}^{\ast}\right\rangle =\left\langle N_{j}^{\ast}\right\rangle =\left\langle N_{i}\right\rangle /\phi$
and $\left\langle N_{i}^{\ast}N_{j}^{\ast}\right\rangle =\left\langle N_{i}^{\ast}\right\rangle ^{2}+O\left(1/S\right)=\left\langle N\right\rangle ^{2}/\phi^{2}$
with $\left\langle N_{i}\right\rangle =\sigma h+\mu$ from Eq. (\ref{eq:LV_to_RE_dictionary})
can be used.

Finally, the distribution of a single element $a_{12}$ can be readily
derived from Eq. (\ref{eq:Pa12a21_of_n}). Integrating over $a_{12}$
weighted by its distribution $P\left(a_{12}\right)$, and using $\left\langle a_{12}\right\rangle =0$,
and then over $n_{1,2}^{\ast}>0$ and using $\dot{\left\langle n_{i}^{\ast}\right\rangle =1/\phi}$
one finds $\Pr\left(a_{12}^{\ast}=a\right)/\Pr\left(a_{12}=a\right)=1-c\cdot a$,
with
\[
c=\frac{\hat{u}/\varphi-h}{\left(q+\sigma_{\lambda}^{2}\right)\varphi}\ .
\]
This equation also holds for $\alpha_{ij},\alpha_{ij}^{\ast}$ since
$\Pr\left(a_{12}^{\ast}=a\right)/\Pr\left(a_{12}=a\right)=\Pr\left(\alpha_{12}^{\ast}=\alpha\right)/\Pr\left(\alpha_{12}=\alpha\right)$
when $\alpha,a$ are related as usual by Eq. (\ref{eq:alpha_a_def}).

\subsection{Two-element distributions\label{appendix:two_elements}}

In this section the joint distribution of two elements, $\alpha_{ij}^{\ast}$
and $\alpha_{kl}^{\ast}$ is calculated. The correlation of $\alpha_{ij}^{\ast}$
with $\alpha_{ji}^{\ast}$ is (to lowest order) the symmetry parameter
$\gamma$, as was shown in the previous section. To order $\,1/S^{2}$,
the only other non-zero correlations are along rows or columns of
the matrix, i.e., when the pairs $\left(i,j\right)$ and $\left(k,l\right)$
share a single index. 
\begin{align}
\hat{u}n_{1}^{+} & =\tilde{h}_{1}-a_{12}n_{2}^{+}-a_{13}n_{3}^{+}\nonumber \\
\hat{u}n_{2}^{+} & =\tilde{h}_{2}-a_{21}n_{1}^{+}-a_{23}n_{3}^{+}\nonumber \\
\hat{u}n_{3}^{+} & =\tilde{h}_{3}-a_{31}n_{1}^{+}-a_{32}n_{2}^{+}\label{eq:np_123_of_h}
\end{align}
The list of the 6 interactions appearing will be denoted by $\left\{ a_{ij}\right\} $,
and the list $\left\{ n_{1}^{+},n_{2}^{+},n_{3}^{+}\right\} $ by
$\left\{ n_{i}^{+}\right\} $. As before, the goal is to calculate
the conditional $P\left(\left\{ a_{ij}\right\} |\left\{ n_{i}\right\} \right)$,
and the same path is followed: first, the joint distribution 
\[
P\left(\left\{ a_{ij}\right\} ,\left\{ n_{i}^{+}\right\} \right)=JP\left(\left\{ a_{ij}\right\} \right)P\left(\tilde{h}_{1}\right)P\left(\tilde{h}_{2}\right)P\left(\tilde{h}_{3}\right)
\]
is calculated, where the Jacobian is 
\[
J=\frac{1}{\hat{u}}-\frac{a_{12}a_{21}+a_{13}a_{31}+a_{23}a_{32}}{\hat{u}^{3}}
\]
and $\tilde{h}_{1,2,3}$ are substituted by their values from Eq.
(\ref{eq:np_123_of_h}). Expanding $P\left(\tilde{h}_{1}\right)$
to second order in $a_{ij}$, 

\begin{multline*}
P\left(\tilde{h}_{1}\right)=\\
P\left(n_{1}^{+}\right)\left[1-\frac{h-\hat{u}n_{1}^{+}}{q+\sigma_{\lambda}^{2}}\omega_{123}+\frac{(h-\hat{u}n_{1}^{+})^{2}-q-\sigma_{\lambda}^{2}}{2\left(q+\sigma_{\lambda}^{2}\right)^{2}}\omega_{123}^{2}\right]
\end{multline*}
where $\omega_{123}\equiv a_{12}n_{2}^{+}+a_{13}n_{3}^{+}$. Expanding
$P\left(\left\{ a_{ij}\right\} |\left\{ n_{i}^{+}\right\} \right)\propto P\left(\left\{ a_{ij}\right\} ,\left\{ n_{i}^{+}\right\} \right)$
to $O\left(a_{ij}^{2}\right)$,
\[
P\left(\left\{ a_{ij}\right\} |\left\{ n_{i}^{+}\right\} \right)\propto P\left(\left\{ a_{ij}\right\} \right)\left[1+\left(..\right)\right]\ .
\]
where the terms $\left(..\right)$\ in the brackets are first and
second powers of $\left\{ a_{ij}\right\} $. The different moments
can now be calculated, remembering to divide by the normalization
that is not trivial to $O\left(a_{ij}^{2}\right)$. The cross-correlations
are
\begin{align*}
\corr_{_{n_{1,2,3}^{\ast}}}\left(a_{12},a_{13}\right) & =-\frac{n_{2}^{\ast}n_{3}^{\ast}}{S^{2}\left(q+\sigma_{\lambda}^{2}\right)}\\
\corr_{_{n_{1,2,3}^{\ast}}}\left(a_{12},a_{31}\right) & =-\frac{\gamma n_{2}^{\ast}n_{3}^{\ast}}{S^{2}\left(q+\sigma_{\lambda}^{2}\right)}\\
\corr_{_{n_{1,2,3}^{\ast}}}\left(a_{21},a_{31}\right) & =-\frac{\gamma^{2}n_{2}^{\ast}n_{3}^{\ast}}{S^{2}\left(q+\sigma_{\lambda}^{2}\right)}
\end{align*}
These results require that the third moments $\mu_{3}=\left\langle a_{ij}^{3}\right\rangle $
decay faster than $O\left(1/S\right)$, since they generate a correction
of order $\mu_{3}/S$. This is rather mild: if one rescales a given
distribution, $P\left(a_{ij}\right)=Sf\left(Sa_{ij}\right)$, then
$\mu_{3}=O\left(S^{-3/2}\right)$.

Going back to variables $\alpha_{ij}$ and $N_{i}$, $n_{2}^{\ast}n_{3}^{\ast}=N_{2}^{\ast}N_{3}^{\ast}/\left\langle N\right\rangle ^{2}$
and $q+\sigma_{\lambda}^{2}=\left\langle N^{2}\right\rangle /\left\langle N\right\rangle ^{2}+\sigma_{K}^{2}/\sigma^{2}$.
For $\sigma_{K}=0$ these become the relations in and following Eq.
\ref{eq:mean_shift}. The correlations over $\alpha^{\ast}$ shown
in Fig. \ref{fig:two_alph_corr}, are obtained by integrating the
above relations over $n_{1,2,3}>0$. For example, 
\begin{align*}
\left\langle \alpha_{12}^{\ast}\alpha_{23}^{\ast}\right\rangle -\left\langle \alpha_{12}^{\ast}\right\rangle ^{2} & =\sigma^{2}\left(\left\langle a_{12}^{\ast}a_{23}^{\ast}\right\rangle -\left\langle a_{12}^{\ast}\right\rangle ^{2}\right)\\
 & =-\frac{\sigma^{2}}{S^{2}\left(q+\sigma_{\lambda}^{2}\right)}\left\langle n_{2}^{\ast}n_{3}^{\ast}\right\rangle -\left\langle \alpha_{12}^{\ast}\right\rangle ^{2}
\end{align*}
where $\left\langle \alpha_{12}^{\ast}\right\rangle $ is given in
Eq. (\ref{eq:avg_alph}), $\left\langle n_{2}^{\ast}n_{3}^{\ast}\right\rangle =\left\langle n_{i}^{\ast}\right\rangle ^{2}=1/\phi^{2}$,
and $q+\sigma_{\lambda}^{2}=\left\langle n^{2}\right\rangle +\sigma_{K}^{2}/\sigma^{2}$.

\subsection{Correlations of interspecies interactions and carrying capacities}

The interactions $\alpha^{\ast}$ and the vector of carrying capacities
of persistent $\vec{K}^{\ast}$ become correlated. The derivation
of these correlations is very similar to the ones in Sec. \ref{appendix:carry_capac},\ref{appendix:two_elements}
above, and is only sketched.

Two additional species are introduced, with
\begin{align}
\hat{u}n_{1}^{+} & =\lambda_{1}+\eta_{1}-a_{12}n_{2}^{+}+h\nonumber \\
\hat{u}n_{2}^{+} & =\lambda_{2}+\eta_{2}-a_{21}n_{1}^{+}+h\label{eq:lam_a_corr_cavity}
\end{align}
where $\eta_{i}\equiv-\sum_{j}a_{ij}n_{j\backslash i}$. The joint
distribution $P\left(\lambda_{1,2},a_{12,21},n_{1,2}^{+}\right)$
is given by 
\[
JP\left(a_{12,21}\right)P\left(\lambda_{1,2}\right)P\left(\eta_{1,2}\right)
\]
where the Jacobian is $J=\left\vert \partial\eta_{1,2}/\partial n_{1,2}^{+}\right\vert $.
$\eta_{1,2}$ are substituted from Eq. (\ref{eq:lam_a_corr_cavity}),
and $P\left(\eta_{1}\right)=g\left(\eta_{1};\lambda_{1}+h,q\right)$
is expanded to first order in $a_{12,21}$. The conditional distribution
is $P\left(\lambda_{1,2},a_{12,21}|n_{1,2}^{+}\right)=CP\left(\lambda_{1,2},a_{12,21},n_{1,2}^{+}\right)$,
where the prefactor $C$ depends on $n_{1,2}^{+}$. The moments of
$P\left(\lambda_{1,2},a_{12,21}|n_{1,2}^{+}\right)$ can now be calculated.
As in previous sections, when $n_{1},n_{2}>0$, then $n_{i}^{+}=n_{i}^{\ast}$.
The new covariance elements read
\begin{align*}
\Sigma_{a_{12},\lambda_{1}} & =\frac{n_{2}^{\ast}}{S\left(1+q/\sigma_{\lambda}^{2}\right)}\\
\Sigma_{a_{12},\lambda_{2}} & =\frac{\gamma n_{1}^{\ast}}{S\left(1+q/\sigma_{\lambda}^{2}\right)}
\end{align*}

\begin{figure}[ptb]
\centering{}\includegraphics[width=1\columnwidth]{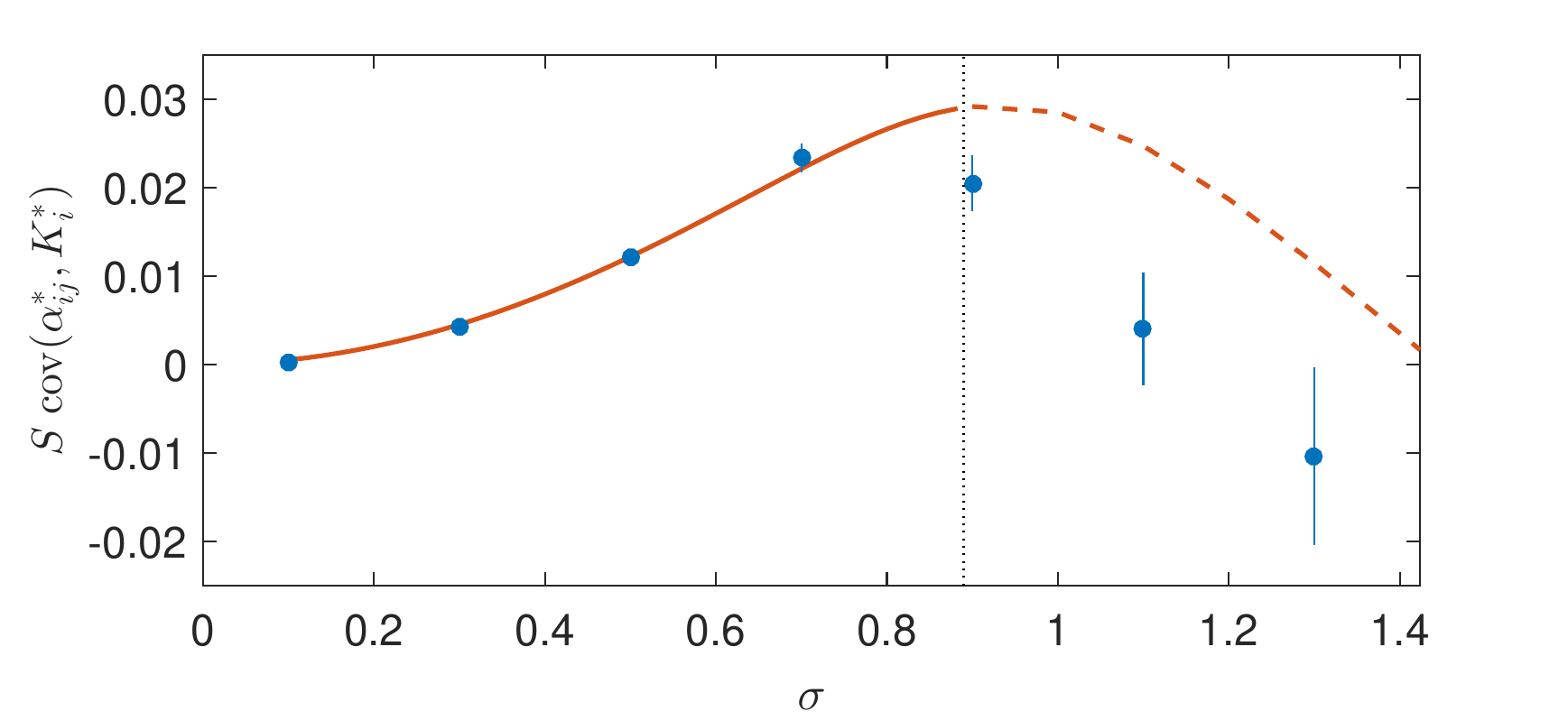}\caption{Correlations of carrying capacities and interspecies interactions
similar to Fig. \ref{fig:carrying_capacity}(c), for $\gamma=1$ (symmetric
$\alpha^{\ast}$), $\mu=8$ and $\sigma_{k}=0.2$.}
\label{fig:carrying_capacity_gam_1}
\end{figure}

Correlations between $\lambda_{1,2}$ and $a_{12,21}$ with no reference
to the abundances, are obtained by integrating over $n_{1,2}^{\ast}$.
The moment $\left\langle \lambda_{1}^{\ast}a_{12}^{\ast}\right\rangle =\left\langle \Sigma_{a_{12},\lambda_{1}}\right\rangle +\left\langle \bar{\lambda}_{1}\mean_{n_{1,2}^{\ast}}a_{12}\right\rangle $,
where $\bar{\lambda}_{1}$ is the mean of $\lambda_{1}$ at given
$n_{1}^{\ast}$, see Eq. (\ref{eq:lam_given_n}) and $\mean_{n_{1,2}^{\ast}}a_{12}$
is given in Eq. (\ref{eq:mu_12}). The covariance $\cov\left(a_{12}^{\ast},\lambda_{1}^{\ast}\right)$
reads
\[
\left\langle a_{12}^{\ast}\lambda_{1}^{\ast}\right\rangle -\left\langle \mu_{12}\right\rangle \left\langle \bar{\lambda}_{1}\right\rangle =\frac{\hat{u}(1+\gamma)\left\langle n_{i}^{\ast}\right\rangle -h\gamma}{S\left(1+q/\sigma_{\lambda}^{2}\right)}\hat{u}\var\left(n_{i}^{\ast}\right)\ .
\]
And one may use $\left\langle n_{i}^{\ast}\right\rangle =1/\phi$
and $\left\langle \left(n_{i}^{\ast}\right)^{2}\right\rangle =q/\phi$
to relate these to the model parameters. Similarly, $\cov\left(a_{12}^{\ast},\lambda_{2}^{\ast}\right)$
is given by
\begin{align*}
 & \left\langle \lambda_{2}^{\ast}a_{12}^{\ast}\right\rangle -\left\langle \bar{\lambda}_{2}\right\rangle \left\langle \mu_{12}\right\rangle =\\
 & \frac{\sigma_{\lambda}^{2}\left\langle n_{i}^{\ast}\right\rangle }{S\sigma_{\lambda}^{2}\left(1+q/\sigma_{\lambda}^{2}\right)^{2}}\left[\begin{array}{c}
\left(\sigma_{\lambda}^{2}+q\right)\gamma\left\langle n_{i}^{\ast}\right\rangle -\hat{u}h\var\left(n_{i}^{\ast}\right)\\
-\gamma\hat{u}h\left(\left\langle \left(n_{i}^{\ast}\right)^{2}\right\rangle +q\phi\left\langle n_{i}^{\ast}\right\rangle ^{2}\right)
\end{array}\right]\ .
\end{align*}
These are translated to $\cov\left(\alpha_{12}^{\ast},K_{1}^{\ast}\right)$
and $\cov\left(\alpha_{12}^{\ast},K_{2}^{\ast}\right)$ plotted in
Fig. \ref{fig:carrying_capacity} by using Eqs. (\ref{eq:alpha_a_def},\ref{eq:LV_to_RE_dictionary}).
Fig. \ref{fig:carrying_capacity_gam_1} shows the covariance for $\gamma=1$
(here $\alpha^{\ast}$ is symmetric so $\cov\left(\alpha_{12}^{\ast},K_{1}^{\ast}\right)=\cov\left(\alpha_{12}^{\ast},K_{2}^{\ast}\right)$).
The analytical results predict that this correlation will be positive.
In the second phase, the analytical predictions are approximate, and
numerics show a transition to negative correlations.

\section{Phase diagram\label{appendix:phase_diag}}

\begin{figure}[ptb]
\centering{}\includegraphics[width=1\columnwidth]{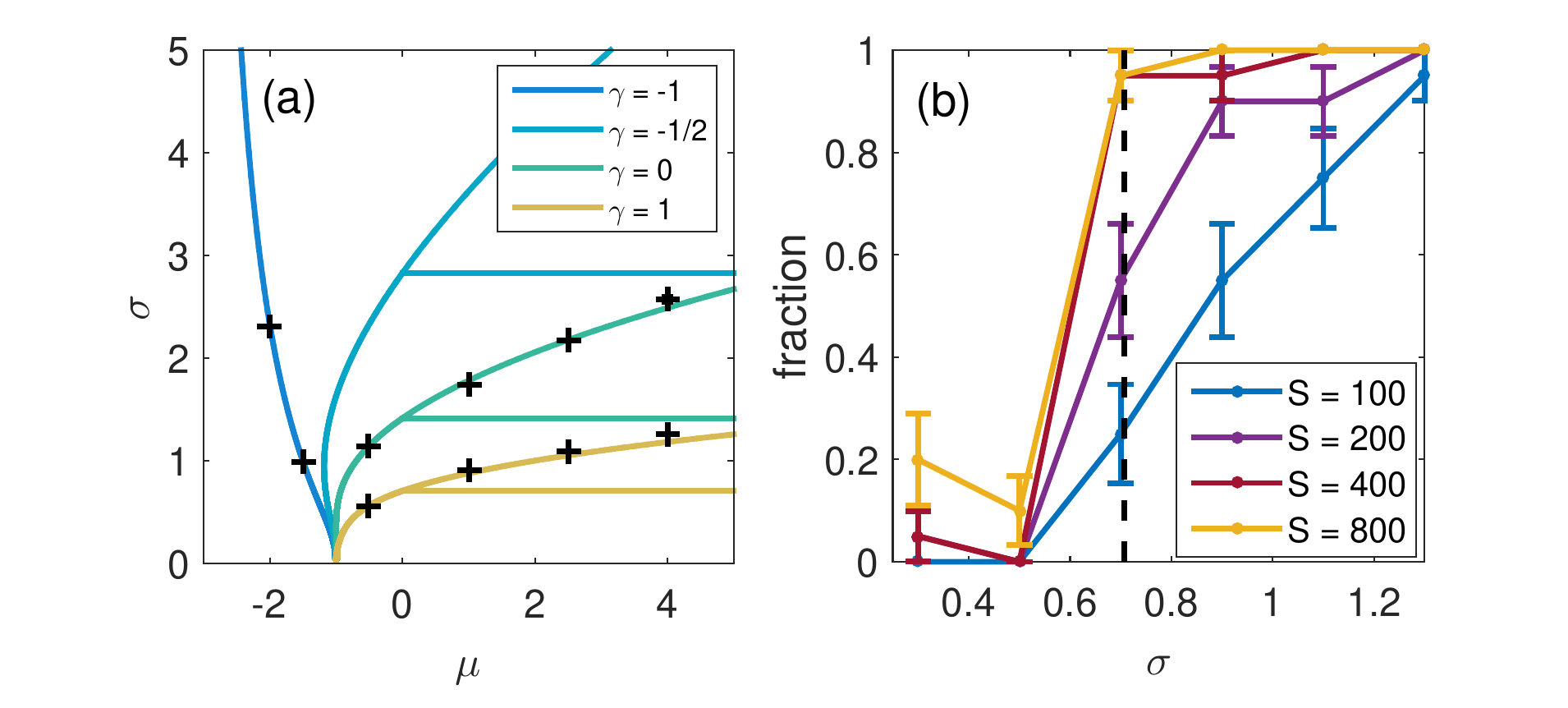}\caption{(a) Phase-diagram for $\sigma_{K}=0$ and different $\gamma$. The
$\gamma=0$ lines correspond to Fig. \ref{fig:phase_diagram_gam_0}
in the main text. Crosses mark transition to diverging solutions,
found numerically. (b) Numerical check of the phase boundary between
first and second phases. The fraction of systems ($\alpha^{\ast}$)
for which there are multiple solutions, at $\mu=4$ and $\gamma=1$.
At large $S$ this fraction jumps sharply at the phase-transition.
The dashed line marks the analytically calculated transition point,
$\sigma=1/\sqrt{2}$.}
\label{fig:phase_diagrams_v2}
\end{figure}

\begin{figure}[ptb]
\centering{}\includegraphics[width=0.7\columnwidth]{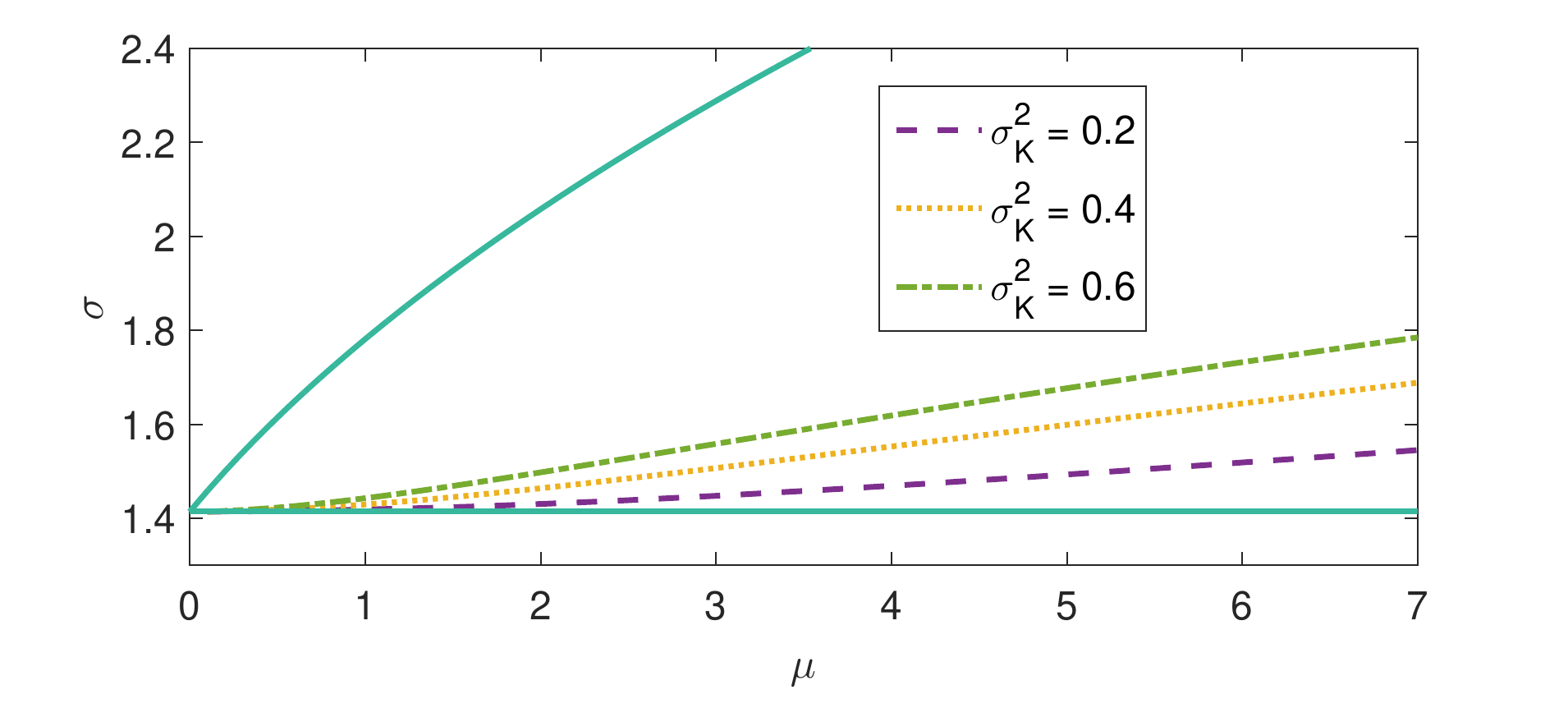}\caption{Phase-boundary between first and second phase for different $\sigma_{K}$,
at $\gamma=0$. Solid lines correspond to the $\gamma=0$ phase boundaries
in Fig. \ref{fig:phase_diagrams_v2}(a).}
\label{fig:phase_diagram_var_car_cap}
\end{figure}

Depending on the parameters $\mu,\sigma,\sigma_{K}$ and $\gamma$,
the model exhibits three distinctive phases, which at large $S$ are
separated by sharp boundaries, see Figs. \ref{fig:phase_diagrams_v2},\ref{fig:phase_diagram_var_car_cap}.
In the first phase, a given system admits a unique equilibrium solution
that is resistant to invasion. In the second phase multiple dynamical
attractors generally exist, which may be stable equilibria or other
attractors such as limit cycles, and the community composition depends
assembly history. In this phase an uninvadable state might not be
reached, and instead invasions trigger jumps between a number of possible
communities \citep{law_permanence_1996,bastolla_biodiversity_2005,capitan_statistical_2009}.
This may happen for dynamical attractors \citep{law_permanence_1996},
or if species that go below some abundance cut-off are removed from
the community, as seems inevitable in any realistic situation \citep{law_permanence_1996,bastolla_biodiversity_2005,capitan_statistical_2009}.
In the present model, we only find it in the second phase, and only
for asymmetric models (e.g. $\gamma=0$). This is further discussed
in the context of the numerical simulations, Appendix \ref{appendix:numerics}.
The transition between the first and second phase is closely related
to those found in various models \citep{diederich_replicators_1989,opper_phase_1992,galla_dynamics_2005,galla_random_2006,yoshino_statistical_2007,yoshino_rank_2008,fisher_transition_2014},
and is also similar to a transition described in \citep{kessler_generalized_2015}.
Finally, in the third phase the abundances grow without bound. At
smaller values of $S$ the transitions between different regimes is
smooth. In particular, for small $S$ the first phase extends further,
as smaller systems have a larger probability to have a unique equilibrium.

The position of the transition to the unbounded growth phase can be
calculated by asking where $\left\langle N\right\rangle $ diverges.
Using the theoretical tools presented in Appendix \ref{appendix:derivations},
by Eq. (\ref{eq:LV_to_RE_dictionary}), $\left\langle N\right\rangle =1/\left(\sigma h+\mu\right)$
so the boundary with the unbounded growth phase lies on the line $\sigma h+\mu=0$.
$h$ is a known function defined in Appendix \ref{appendix:derivations},
following Eq. (\ref{eq:order_param_v_eqn}). The analytical expression
for $h$ is exact in the first phase and approximate in the second,
so the prediction for this phase boundary will accordingly be exact
when it limits the first phase, and approximate when it limits the
second.

The boundary between the first and second phases lies on the line
$\phi=\left(u-\gamma v\right)^{2}$, where $\phi$ is the fraction
of persistent species, and $v$ is a known function, see Appendix
\ref{appendix:derivations}. For $\sigma_{K}^{2}=0$ this line lies
at $\sigma=\sqrt{2}/\left(1+\gamma\right)$ for all $\mu>0$. Along
this line the linear response of a the abundances to a change in the
carrying capacities diverges, indicating loss of stability of the
unique equilibrium solution and the appearance of multiple attractors.
More precisely, the change of the normalized abundances $\vec{n}$
in response to a perturbation $\vec{\xi}$ defined in Eq. (\ref{eq:RE_version_with_response})
is $\left\langle \left(\delta n\right)^{2}\right\rangle /\left\langle \xi^{2}\right\rangle =\phi/\left[\left(u-\gamma v\right)^{2}-\phi\right]$,
when the $\xi_{i}$'s are sampled independently (the average includes
$\delta n_{i}=0$ for species outside the community). This transition
line can be derived using known techniques \citep{opper_phase_1992}
similar to the arguments in Appendix \ref{appendix:derivations}.
This phase transition is only encountered when the average interaction
is competitive, i.e. for $\mu>0$ and therefore could not be seen
in \citep{rieger_solvable_1989}, where a Lotka-Volterra system was
studied with $\mu=0$.

\section{Small $S$\label{appendix:small_S}}

\begin{figure}
\centering{}\includegraphics[width=1\columnwidth]{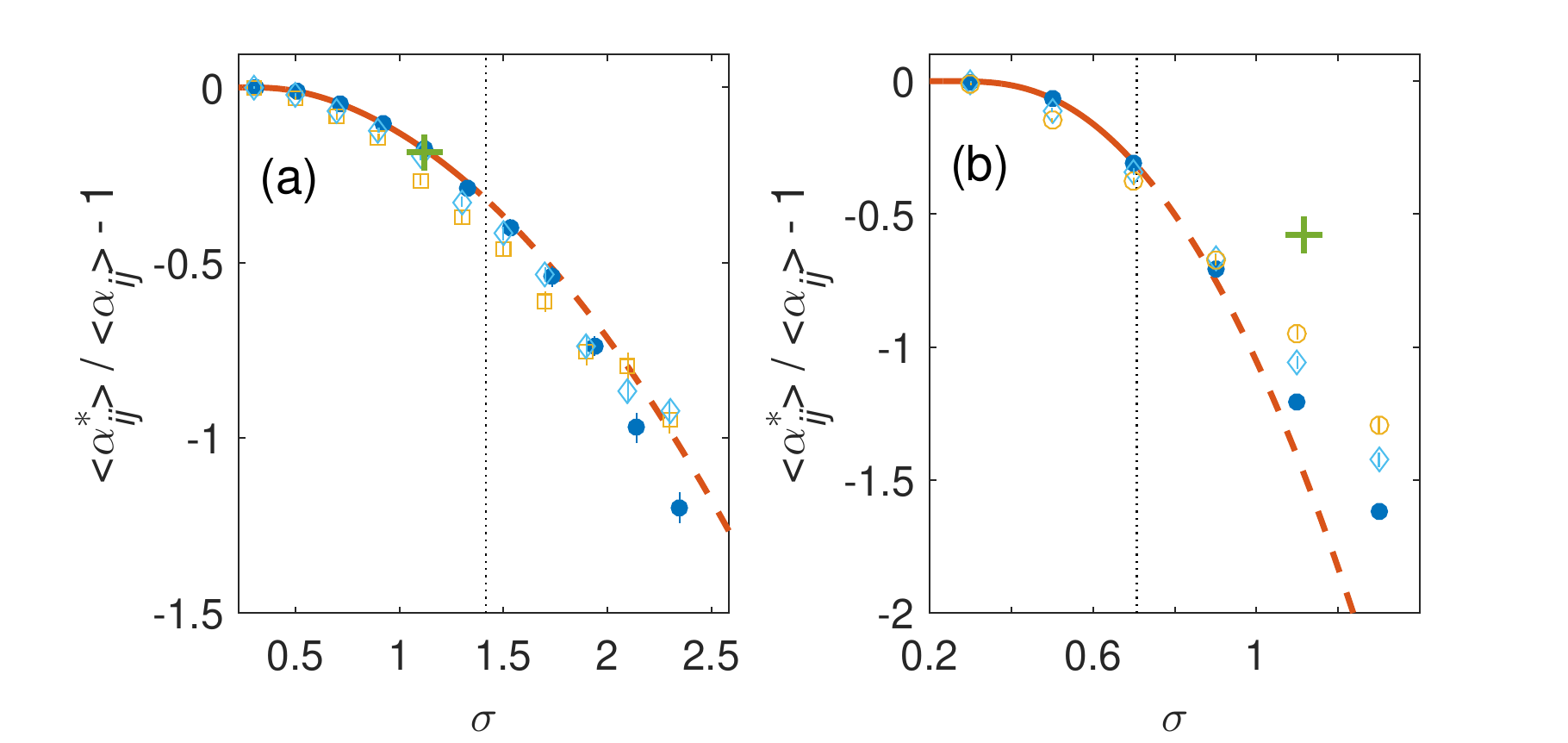}\caption{(a,b) are the same as Fig. \ref{fig:alpha_mean_and_distrib}(c,d)
respectively, with additional numerical results. Diamonds: $\alpha_{ij}$
sampled from a Gaussian distribution, $S=25$. Crosses: $S=15$ with
$\alpha_{ij}$ sampled from a uniform distribution on $\left[0,1\right]$.}
\label{fig:alpha_mean_small_S}
\end{figure}

\begin{figure}
\centering{}\includegraphics[width=1\columnwidth]{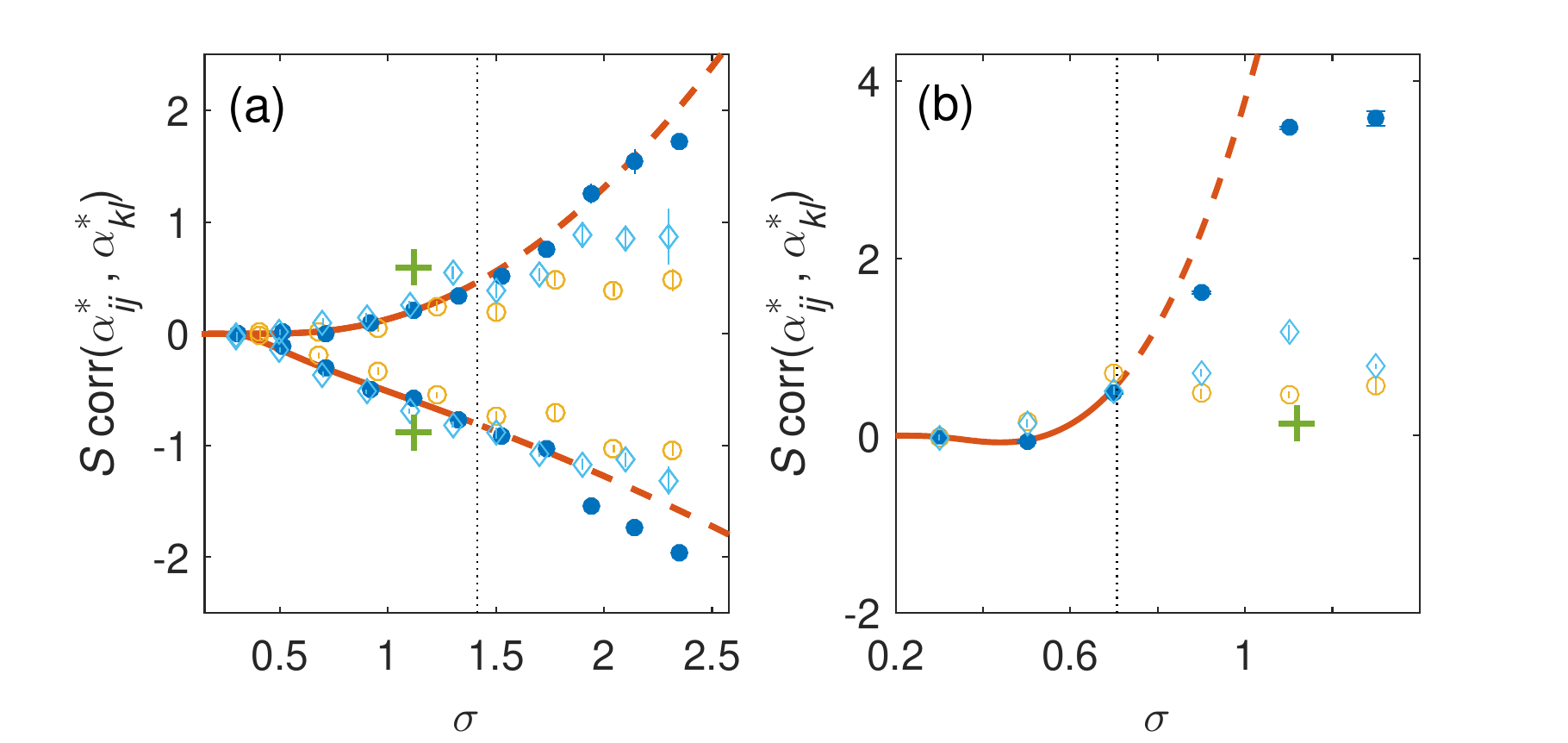}\caption{Same as Fig. \ref{fig:alpha_mean_and_distrib}(c,d) respectively,
with additional numerical results. Symbols as in Fig. \ref{fig:alpha_mean_small_S}.}
\label{fig:two_alph_corr_small_S}
\end{figure}

Figs. \ref{fig:alpha_mean_small_S},\ref{fig:two_alph_corr_small_S}
are identical to Fig. \ref{fig:alpha_mean_and_distrib}(a,b) and Fig.
\ref{fig:two_alph_corr}, with additional simulations for small pool
sizes. As in the main text, the numerical results are plotted as a
function of $\sigma=\sqrt{S}\std\left(\alpha_{ij}\right)$, and the
analytical predictions as a function of $1/u\simeq\sigma$, see Eq.
(\ref{eq:LV_to_RE_dictionary}). For normally distributed $\alpha_{ij}$,
numerics for pools of size $S=15,25$ are shown in addition to the
$S=200$ results. For $S=15$, $\alpha_{ij}$ is the mean of is 0.27
and the standard deviation up to 0.6. Another comparison is with $\alpha_{ij}$
sampled from uniform distribution on $\left[0,1\right]$, with $S=15$
and community sizes of about 6-7 species. The results are in good
agreement with numerics even for the $S=15$ numerics, in the region
were the analytics are exact (unique equilibrium phase, left of dotted
vertical line).

Depending on the application, one might wish to study models where
interactions are purely competitive, or a combination of competitive
and beneficial interactions. In addition to the choice of the distribution,
the combination of $S,\mu$ and $\sigma$ at a given (e.g., Gaussian)
distribution allows for similar control. The fraction of beneficial
interactions ($\alpha_{ij}<0$) is given by the area of the negative
tail of $P\left(\alpha_{ij}\right)$. For $S=15$ in Figs. Figs. \ref{fig:alpha_mean_small_S},\ref{fig:two_alph_corr_small_S},
at $\sigma=0.5$ only about 2\% of the interactions will be beneficial,
and only mildly so. Below $\sigma=0.4$, typically only one or less
of the interactions will be beneficial. At larger widths the $\alpha_{ij}$
combine competitive and beneficial interactions.

\section{Numerical simulations\label{appendix:numerics}}

To numerically find persistent solutions, the network variables $\alpha_{ij}$
and $K_{i}$ are first sampled. $\alpha_{ij}$ are sampled from a
normal distribution unless otherwise stated. A uniform distribution
was checked to give identical results at large $S$. Results for small
$S$ are given in Sec. \ref{appendix:small_S}.

The Lotka-Volterra dynamics, Eq. (\ref{eq:LV}), are then integrated
using a Runga-Kutta 45 solver, from random initial conditions sampled
uniformly on $\left[0,1\right]$. All species that go below an abundance
cut-off $N_{i}<10^{-14}$ are removed from the community ($N_{i}$
set to zero). The solver is terminated when an equilibrium solution
is found, in which for every $i$ either $dN_{i}/dt$ is small, or
$N_{i}<10^{-14}$. Solutions that do not terminate are stopped after
a long time ($T=10^{7}$) and all variables with $N_{i}>10^{-14}$
are considered part of the community. The solution is checked against
invasion of the pool species not present. As some species are removed
during the dynamics due to the abundance cut-off, it is possible that
they would be able to invade later. If any such species are found,
the dynamics are run from the end point of the first simulation with
additional small abundance ($N_{i}=10^{-10}$) to the species that
may invade. This process is repeated until an uninvadable solution
is reached or after ten iterations. In phase one the resulting community
is always found to be uninvadable, and usually reached on the first
run of the dynamics. For a given system all initial conditions give
the same final community. In phase two for asymmetric interactions
(specifically $\gamma=0$), this process did not always converge to
an uninvadable solution after ten iterations and then was stopped.
All numerical results shown in the paper show only minor differences
when plotted after the first run, compared to iterations of the invasion
process.

\begin{figure}[H]
\begin{centering}
\includegraphics[width=1\columnwidth]{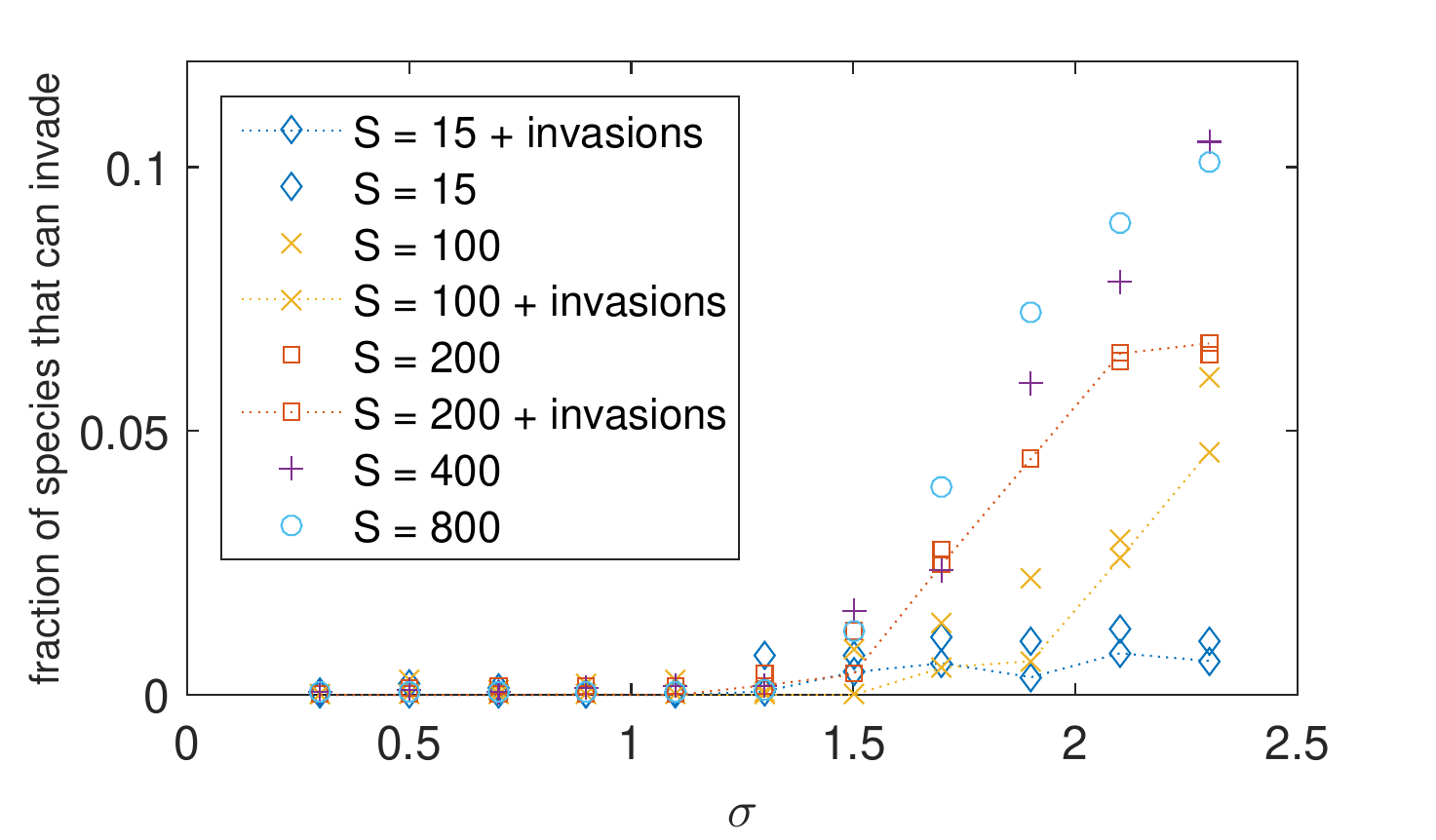}
\par\end{centering}
\caption{Fraction of species that may invade. Data plotted with dashed lines
show the results after multiple invasion attempts.}

\label{fig:invasion tests}
\end{figure}

Results large $S$ were simulated with $S=400$. An exception are
the results for $\gamma=0$ in Figs. \ref{fig:alpha_mean_and_distrib},\ref{fig:two_alph_corr}
which were run with $S=200$. This was chosen as balance finite-size
effects while minimizing the number of species that can invade in
phase two: The results for $S=100$ and $S=200$ are very similar,
indicating good finite-size convergence, and both have a small fraction
(less than $0.05$, see Fig. \ref{fig:invasion tests}). Other options
are possible, and would represent ecological conditions with varying
effects of the minimal allowed abundance.

To test for multiplicity of equilibria, as shown in Fig. \ref{fig:phase_diagrams_v2}(b),
the same system (same $\alpha_{ij}$ and $K_{i}$) is run starting
from different initial conditions.

\end{document}